\documentclass[twocolumn,aps,prd,preprintnumbers,superscriptaddress,nofootinbib,10pt]{revtex4-1}
\usepackage{epsfig}
\usepackage{bm}
\usepackage{amssymb}
\usepackage{amsmath}
\usepackage{color}
\usepackage{subfigure}
\usepackage[colorlinks,
            linkcolor=blue,
            anchorcolor=black,
            citecolor=blue
            ]{hyperref}

\newcommand{\beq}{\begin{eqnarray}}
\newcommand{\eeq}{\end{eqnarray}}

\newcommand{\nn}{\nonumber \\}

\begin{document}
\title{Azimuthal Angular Asymmetry of Soft Gluon Radiation in Jet Production  }

\author{Yoshitaka Hatta}
\affiliation{Physics Department, Building 510A, Brookhaven National Laboratory, Upton, NY 11973, USA}
\affiliation{RIKEN BNL Research Center,  Brookhaven National Laboratory, Upton, NY 11973, USA}

\author{Bo-Wen Xiao}
\affiliation{School of Science and Engineering, The Chinese University of Hong Kong, Shenzhen 518172, China}

\author{Feng Yuan}
\affiliation{Nuclear Science Division, Lawrence Berkeley National
Laboratory, Berkeley, CA 94720, USA}

\author{Jian Zhou}
\affiliation{\normalsize\it  Key Laboratory of Particle Physics and
Particle Irradiation (MOE),Institute of Frontier and
Interdisciplinary Science, Shandong University (QingDao), Shandong
266237, China }


\begin{abstract}
We investigate the impact of soft gluon resummation on the azimuthal angle correlation between the total and relative momenta of two energetic final state particles (jets). We show that the initial and final state radiations induce sizable  $\cos(\phi)$ and $\cos (2\phi)$  asymmetries in single jet and dijet events, respectively. We numerically evaluate the magnitude of these asymmetries for a number of processes in collider experiments, including diffractive dijet and dilepton production in ultraperipheral $pA$ and $AA$ collisions, inclusive and diffractive dijet production at the EIC and inclusive dijet production in $pp$ collisions at the LHC. In particular, the $\cos (2\phi)$ asymmetry of perturbative origin can dominate over the primordial asymmetry due to the linearly polarized gluon distribution.   
\end{abstract}
\maketitle

\section{Introduction}

\begin{figure}[tbp]
\begin{center}
\includegraphics[width=0.4\textwidth]{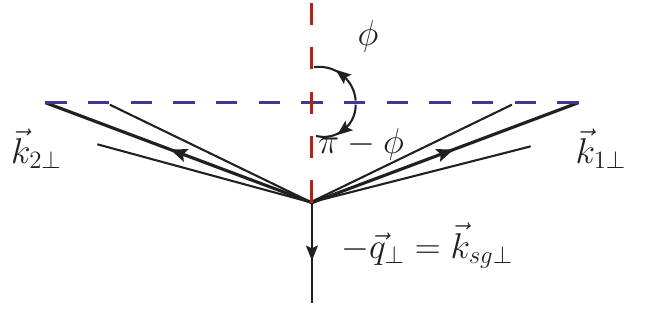}
\end{center}
\caption[*]{Dijet in transverse plane perpendicular to the beam direction at hadron colliders. The dijet total transverse momentum $\vec{q}_\perp=\vec{k}_{1\perp}+\vec{k}_{2\perp}$, which is due to multiple soft gluon emissions, is much smaller than the individual jet momentum $\vec{P}_\perp=(\vec{k}_{1\perp}-\vec{k}_{2\perp})/2$.   } 
\label{fig:dijet0}
\end{figure}

Jet processes are the most abundant events in hard hadronic collisions and have been under intensive investigations at various colliders, see, e.g., Refs.~\cite{Abazov:2004hm,Abelev:2007ii,Khachatryan:2011zj,daCosta:2011ni,Aad:2010bu,Chatrchyan:2011sx,Adamczyk:2013jei,Aaboud:2019oop}. Among them, the productions of dijet and jet plus a color-neutral particle (such as the Higgs boson) are characterized by distinct final states where two energetic particles (jets) are  almost back-to-back in the transverse plane perpendicular to the beam direction. Deviations from the exactly back-to-back configuration are in general expected. In other words,  the total transverse momentum of the two outgoing systems  $\vec{q}_\perp=\vec{k}_{1\perp}+\vec{k}_{2\perp}$ is typically small but nonzero, as illustrated in Fig.~\ref{fig:dijet0} for the dijet case. The cross section then depends on the angle $\phi$ between $\vec{q}_\perp$ and the dijet relative momentum $\vec{P}_\perp=(\vec{k}_{1\perp}-\vec{k}_{2\perp})/2$  
\begin{equation}
\frac{d\sigma}{dP_\perp dq_\perp d\phi} = \sigma_0 + \cos (\phi) \sigma_1+\cos(2\phi) \sigma_2+\cdots.
\end{equation}
The coefficients $\sigma_1,\, \sigma_2,\,\cdots$ often encode novel partonic structures  of the target that are important in the study of nucleon tomography at the future electron-ion collider (EIC)~\cite{Accardi:2012qut,Proceedings:2020eah, AbdulKhalek:2021gbh}. A primary example is exclusive diffractive dijet production in $\gamma^{(*)}p$ scattering where $q_\perp$ is provided by the recoil momentum of the target. It has been predicted that the `elliptic' gluon Wigner distribution  generates a $\cos (2\phi)$ asymmetry   \cite{Hatta:2016dxp,Altinoluk:2015dpi,Zhou:2016rnt,Hagiwara:2017fye,Mantysaari:2019csc,Mantysaari:2019hkq}. Another example is the inclusive dijet production in DIS, where $q_\perp$ comes from  the intrinsic transverse momentum of gluons in the target. In this case, the so-called linearly polarized gluon distribution can generate a $\cos (2\phi)$ asymmetry in the dijet system  \cite{Boer:2010zf,Metz:2011wb,Dumitru:2015gaa,Boer:2017xpy,Boer:2016fqd,Xing:2020hwh,Zhao:2021kae}.  

However, the momentum imbalance $q_\perp$ can simply come from perturbative initial and final state radiations which have nothing to do with nontrivial parton distributions inside the target. Depending on kinematics, this can affect or even dominate the coefficients $\sigma_1,\sigma_2,\cdots$ when $P_\perp \gg q_\perp$. The reason is that the radiative corrections are enhanced by large double logarithms 
$(\alpha_s \ln^2 P_\perp^2/q_\perp^2)^n$. While the resummation of these logarithms is well  understood for the angular independent part $\sigma_0$  \cite{Banfi:2003jj,Banfi:2008qs,Hautmann:2008vd,Mueller:2013wwa,Sun:2014gfa,Sun:2015doa,Hatta:2019ixj,Liu:2018trl,Liu:2020dct,Chien:2019gyf,Chien:2020hzh}, that for the angular dependent part has been discussed much less frequently in the literature. In a series of papers by Catani {\it et al.}    \cite{Catani:2014qha,Catani:2017tuc}, it has been observed that the resummation for $\sigma_{1},\sigma_2,\cdots$  can be done in the Fourier space $\vec{q}_\perp\to \vec{b}_\perp$ using the same Sudakov factor as for $\sigma_0$. An interesting new feature is that although the angular dependent cross section is singular $1/q_\perp^2$ in fixed-order calculations with no compensating virtual correction, the resummed cross sections $\sigma_{1,2}$ are well-behaved as $q_\perp\to 0$. The goal of this paper is to study this resummation in detail and make quantitative predictions for azimuthal asymmetries $\langle \cos (n\phi)\rangle$ that can be compared with the existing and future experimental data. We shall consider 
a variety of processes, including dijet productions in diffractive and inclusive processes, and jet plus color-neutral particle production. A brief summary of our results has been published in Ref.~\cite{Hatta:2020bgy}. 

\begin{figure}[tbp]
\begin{center}
\includegraphics[width=7cm]{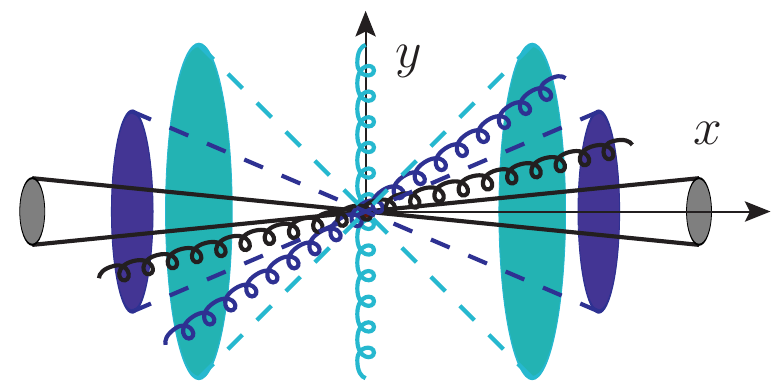}
\end{center}
\caption[*]{An illustration of soft gluon radiations in back-to-back dijet events. Due to the collinear enhancement, soft gluons are more likely emitted closer to jet cones. (The darkness of the gluon color is correlated with the probability of the  emission.) From momentum conservation,  $\vec{q}_\perp=-\sum_i^{soft}\vec{k}_{i\perp}$. Since $\vec{k}_{i\perp}$'s tend to point to jet directions, the same is true for $\vec{q}_\perp$,  resulting in a sizable anisotropy $\langle \cos(2\phi)\rangle$.} 
\label{fig:dijet1}
\end{figure}

An important feature of the final state radiation is that the emitted soft gluons tend to be aligned with jet directions, see Fig.~\ref{fig:dijet1}. Those emitted inside jet cones become part of the jets, so one needs to carefully treat emissions slightly outside the jet cones.  Since $\vec{q}_\perp$ is the recoil momentum against these gluons, it also points towards jet directions on average. This naturally generates a positive $\cos (2\phi)$ asymmetry of purely perturbative origin in dijet events \cite{Hatta:2020bgy}. A recent measurement by the CMS collaboration \cite{CMS:2020ekd} indicates that the magnitude of this effect is sizable. Depending on kinematics, it can completely overshadow the `intrinsic' azimuthal correlations generated by nontrivial parton distributions. For the search of the elliptic Wigner distribution in exclusive dijet production mentioned above, one can avoid this problem by measuring the correlation between $\vec{P}_\perp$ and the nucleon recoil momentum (instead of $\vec{q}_\perp$) as originally suggested in \cite{Hatta:2016dxp}. However, in inclusive dijet production, it is not possible to cleanly separate the contribution from the linearly polarized gluon distribution, unless one has an accurate control of the perturbative backgrounds.

Our discussions in this paper are connected to other recent developments in the field. The correlation between $\vec{q}_\perp$ and $\vec{P}_\perp$ also measures the correlation between final state jets. The combination of this study with three particle correlations in the final state recently proposed in Refs.~\cite{Chen:2020adz,Karlberg:2021kwr,Chen:2021gdk} shall open a new avenue to study the QCD dynamics of gluon radiation. In this regard, the non-global logarithms (NGLs)~\cite{Dasgupta:2002bw,Banfi:2008qs} can also contribute to the observables we consider, although their numerical impact might be limited for the relevant kinematics \cite{Hatta:2019ixj}. More broadly,  the perturbative contribution to the $\cos(2\phi)$ azimuthal asymmetries has been studied for various processes~\cite{Boer:2006eq,Berger:2007si,Bacchetta:2008xw,Bacchetta:2019qkv,Nadolsky:2007ba,Catani:2010pd,Catani:2014qha,Catani:2017tuc,Hatta:2020bgy}. In particular, it may shed light on the QCD factorization and resummation for power corrections in hard scattering processes~\cite{Bacchetta:2019qkv,Balitsky:2017flc,Balitsky:2017gis,Ebert:2018gsn,Moult:2019mog}.

This paper is organized as follows. In Section II, we consider processes in which the final state consists of one jet and one color-neutral particle. In subsection IIA, we consider lepton-jet correlations in $ep$ scattering in the laboratory frame. In subsection IIB, we consider photon-jet production in $pp$ collisions.  The former can be studied at the EIC, whereas the latter can be studied at RHIC and the LHC. 
Since there is only one jet in the final state, we expect that the dominant asymmetry is of the form $\cos (\phi)$. 

In Section III, we study dijet production. In subsections IIIA and IIIB, we consider diffractive and inclusive dijet photo-production processes, respectively. As mentioned above, the dominant asymmetry is $\cos (2\phi)$ in this case. 
Then in subsection IIIC, we consider inclusive dijet production in $pp$ collisions specifically focusing on the (most complicated) $gg\to gg$ channel. 

In Section IV we give a detailed analysis of dijet electro-production in DIS. When the photon is virtual, the linearly polarized gluon distribution gives an additional contribution to the $\cos (2\phi)$ asymmetry. We numerically compare the respective contributions to the asymmetry from nonperturbative and perturbative mechanisms. 

Finally, in Section V, we extend our analysis to  QED processes, where a lepton pair is produced in two photon scattering in ultraperipheral heavy ion collisions (UPCs). Di-lepton production in this process has a long  history~\cite{Bertulani:1987tz,Adams:2004rz,Baur:2003ar,Hencken:2004td,Baur:2007fv,Bertulani:2005ru,Baltz:2007kq,Baltz:2009jk,ATLAS:2016vdy,Klein:2018cjh,CMS:2020avp,Klein:2020fmr} and has attracted great attention recently through comprehensive measurements at RHIC and the LHC~\cite{Aaboud:2018eph,Adam:2018tdm,Lehner:2019amb,Adam:2019mby,ATLAS:2019vxg,Sirunyan:2020vvm,Aad:2020dur}. Theory progress has also been made to understand the underlying physics~\cite{Klusek-Gawenda:2018zfz,Klein:2018fmp,Zha:2018tlq,Li:2019yzy,Li:2019sin,Zhao:2019hta,Karadag:2019gvc,Vidovic:1992ik,Klein:2020jom,Xiao:2020ddm,Klusek-Gawenda:2020eja,Brandenburg:2021lnj}. We will show that the photon radiation can contribute to a significant $\cos(2\phi)$ asymmetries in the kinematic region where the perturbative contribution dominates.

\section{Jet Plus Color-neutral Particle in the Final State}

In this section, we discuss the final states with a jet and a color-neutral particle. The soft gluon radiation comes only from the jet and the incoming parton(s). The dominant azimuthal asymmetry is then the  $\cos(\phi)$ term.  We will first study lepton plus jet production in $ep$ collisions and then extend to photon plus jet production in $pp$ collisions. Similar studies can also be carried out for jet plus Higgs boson or $Z/W$ boson productions at the LHC. 

\subsection{Lepton and Jet Correlation in $ep$ Collisions}

At leading order a lepton scatters off a quark through  virtual photon exchange in $t$-channel and produces a quark jet in the final state
\begin{equation}
    e(k)+q(p_1)\to e'(k_\ell)+jet(k_J)+X \, .
\end{equation}
In the laboratory frame, the final state lepton and jet are mainly back-to-back in the transverse plane perpendicular to the beam direction. This process has recently attracted significant interest because it can provide a novel way to study the transverse momentum dependent (TMD) quark distribution in the nucleon~\cite{Gutierrez-Reyes:2018qez,Gutierrez-Reyes:2019vbx,Liu:2018trl,Arratia:2019vju,Liu:2020dct,Arratia:2020nxw,Kang:2020fka}. It has also motivated experimental efforts to re-analyze the existing HERA data~\cite{Amilkar,Miguel}.

The virtuality of the photon defines the hard-scattering process. To leading order, the differential cross section can be written as
\begin{equation}
\frac{d^5\sigma^{ep\to e'qX}}{dy_\ell d^2 P_{\perp} d^2q_{\perp}}=\sigma_0^{eq} xf_q(x) \delta^{(2)}(q_\perp) \ ,
\end{equation}
where $\sigma_0^{eq}=\frac{\alpha_e^2e_q^2}{\hat s Q^2}\frac{2(\hat s^2+\hat u^2)}{Q^4}$, $y_\ell$ is the rapidity of the final state lepton in the laboratory frame. Following the notations in Introduction, we have define the difference and total transverse momenta for the two final state particles:  $\vec{P}_\perp=(\vec{k}_{\ell\perp}-\vec{k}_{J\perp})/2$ and $\vec{q}_\perp=\vec{k}_{\ell \perp}+\vec{k}_{J\perp}$. [Below we  often omit an arrow on transverse vectors.] In the above equation, $x$ represents the momentum fraction of the incoming nucleon carried by the quark, $f_q(x)$ for the quark distribution function. The Mandelstam variables $\hat s$, $\hat t$ and $\hat u$ are defined as usual for the partonic sub-process, in particular, $\hat t=(k_\ell-k)^2=-Q^2$. 
 
At one-loop order, $q_\perp$ can be nonzero due to the emission of a soft gluon  with momentum $k_{\perp g}$~\cite{Liu:2020dct}. Integration over the phase space of the emitted gluon is explained in Appendix A. The result is 
\begin{eqnarray}
&& g^2\int \frac{d^3k_g}{(2\pi)^32E_{k_g}}\delta^{(2)} (q_\perp+k_{g\perp})
C_FS_g(k_J,p_1) \nonumber \\ 
&& = \frac{\alpha_sC_F}{2\pi^2 q_\perp^2}\Biggl[\ln\frac{Q^2}{q_\perp^2}+\ln\frac{Q^2}{k_{\ell\perp}^2}\nonumber\\
&& 
\qquad +c_0+2c_1\cos(\phi)+2c_2\cos(2\phi) +\cdots\Biggr]\  , \label{int}
\end{eqnarray}
where
\begin{equation}
    S_g(k_J,p_1)=\frac{2k_J\cdot p_1}{k_J\cdot k_gp_1\cdot k_g}\, ,
\end{equation}
and $\phi$ is the azimuthal angle between $q_\perp$ and 
$P_\perp$.  
Note that the coefficients $c_n$ in general depend  on $q_\perp$. But the dependence is power-suppressed
 \begin{equation}
 c_n(q_\perp^2)-c_n(0)={\cal O}\left( (q_\perp/P_\perp)^c\right)\,,
 \end{equation}
where the integer $c$ (usually $c=1$ or $2$) depends on both  $n$ and the process under consideration. In (\ref{int}), we recognize at least two  sources of such power corrections. First, when $q_\perp$ is small but nonvanishing,  the soft gluon rapidity is subject to kinematical constraints  $y_{min}<y_g<y_{max}$ with $|y_{max/min}|\sim \ln P_\perp^2/q_\perp^2$. However, in the actual calculation of $c_n$ below,  it is convenient to integrate over  $-\infty <y_g<\infty$. The difference in $c_n$ caused by  this approximation is power-suppressed.  Second, in the soft emission kernel $S_g$, one approximates $k_{J\perp} =\frac{1}{2}q_\perp-P_\perp \approx -P_\perp$. Again the difference is power-suppressed in $q_\perp/P_\perp$. There are also power corrections from the hard part that can affect azimuthal asymmetries. In this paper, we do not study these corrections systematically (they are in any case beyond the leading TMD factorization formalism), and in most of our calculations below,  we neglect the $q_\perp$-dependence of $c_n$. However, in Section IIIA, we will include part of power corrections for phenomenological reasons.
 
When calculating the Fourier coefficients $c_n$, we need to subtract in the $k_g$-integral the configuration where the soft gluon is emitted inside the jet cone of radius $R$. Namely, we have to impose the constraint 
\begin{equation}
\Delta_{k_gk_J} \equiv (y_g-y_J)^2 +(\phi_g-\phi_J)^2>R^2\,. \label{condition}
\end{equation} 
 As a result, $\{c_n\}$ depend on $R$ rather strongly.  To gain analytical  insights into the coefficients, it is convenient to replace (\ref{condition}) by   
\begin{equation}
k_J\cdot k_g\propto 2(\cosh (y_g-y_J)-\cos (\phi_g-\phi_J))>R^2\,, \label{jetdef2}
\end{equation} 
which is equivalent to (\ref{condition}) when $R\ll 1$. 
\begin{figure}[tbp]
\begin{center}
\includegraphics[width=0.4\textwidth]{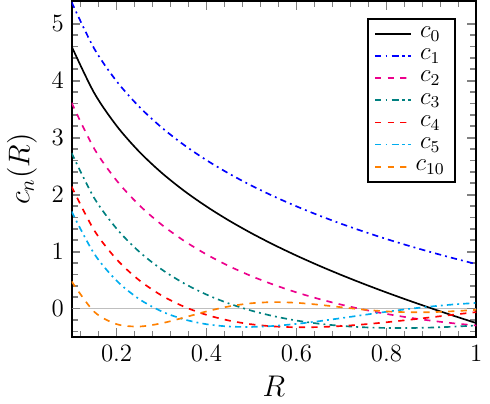}
\end{center}
\caption[*]{Fourier coefficients $c_n (R)$ given by  (\ref{cn}) are shown as a function of $R$.}
\label{fig:fourier}
\end{figure}
We can then obtain the following explicit expression for an arbitrary Fourier coefficient $c_n$ 
 \begin{equation}
     c_n=\ln\frac{1}{R^2} +f(n)+g(nR)\,,
 \end{equation} 
 with
 \begin{eqnarray}
     f(n)&=&\frac{2}{\pi}\int_0^\pi d\phi (\pi-\phi) \frac{\cos \phi }{\sin \phi} \left( \cos n\phi -1\right), \\
     g(nR)&=& \frac{4}{\pi} \int_0^1 \frac{d\phi}{\phi} \tan^{-1} \frac{\sqrt{1-\phi^2}}{\phi} \left[1-\cos \left( nR \phi\right)\right]\nonumber\\
     &=&\frac{n^2R^2}{4} \, _2F_3\left(1,1;2,2,2; \, -\frac{n^2R^2}{4}\right).
 \end{eqnarray}
The collinear singularity is isolated in the logarithm $\ln 1/R^2$, and the remaining part is finite. In particular, $c_0=\ln(1/R^2)$ and the first few coefficients of the rest read $f(1)=2\ln 4-2$, $f(2)=-1$, $f(3)=2\ln 4-14/3$ and $f(4)=-5/2$. For sufficiently large values of $n$, we find  $f(n)\simeq \ln(b_0^2/n^2)$ with $b_0 = 2e^{-\gamma_E}$ ($\gamma_E$ is the Euler constant). Also note that  $g(nR)\approx n^2R^2/4$ when $nR\ll 1$, while $g(nR)\approx \ln (n^2R^2/b_0^2)$ in the limit $nR\gg 1$. This indicates that $c_n$ vanishes when $nR\gg 1$.  

When $R$ is large $\sim {\cal O}(1)$, we should return to (\ref{condition}). The Fourier coefficients can be evaluated numerically as follows (see (\ref{power})) 
\begin{eqnarray}
c_n &=& \frac{2}{\pi}\int_0^R d\phi \frac{\cos \phi }{\sin \phi} \left[(\pi-\phi)  -\tan^{-1} \left( \frac{e^{y_+}-\cos \phi}{\sin \phi}\right)\right.\nonumber\\
&&\left.+\tan^{-1} \left( \frac{e^{y_-}-\cos \phi}{\sin \phi}\right)\right]  \cos n\phi \nonumber \\
  &&+ \frac{2}{\pi}\int^\pi_R d\phi \frac{\cos \phi }{\sin \phi} (\pi-\phi)\cos n\phi \notag \\
 &&  - \frac{2}{\pi}\int_0^R d\phi\, y_+ \cos n\phi, 
  \label{cn}
\end{eqnarray}
where $y_{\pm} = \pm \sqrt{R^2-\phi^2}$. For example, for $R=1$, we have $c_0\simeq -0.25$, $c_1=0.78$ and $c_2=-0.30$. As shown in Fig.~\ref{fig:fourier}, $c_n$ decreases approximately as $\ln 1/R^2$ for small $n$ values, while oscillations around zero start to appear for large-$n$ coefficients.

We now extend the above one-loop results to all orders in the TMD framework by resumming the double and single logarithms in $Q^2/q^2_\perp$. This is appropriately carried out in the Fourier transformed $b_\perp$-space.
The resummed azimuthal averaged cross section reads~\cite{Liu:2020dct},
\begin{eqnarray}
\frac{d^5\sigma^{ep\to e'qX}}{dy_\ell d^2 P_{ \perp} d^2q_{\perp}}&=&\sum_q \sigma_0^{eq}  \int \frac{d^2 b_\perp}{(2\pi)^2}  e^{i q_\perp \cdot b_\perp} x_q  f_q(x_q,\mu_b)  \nonumber \\ 
&& \times   e^{- \text{ Sud}^{eq}(b_\perp,P_\perp,R)} \ ,
\end{eqnarray}
where $\mu_b\equiv b_0/b_\perp$ with $b_0=2e^{-\gamma_E}$ and $\gamma_E$ the Euler constant. Here and in the following, we neglect the high order corrections to the hard factor in the resummation formulas. The Sudakov form factor is defined as
\begin{eqnarray}
 \text{ Sud}^{eq}
 &=& \int^{Q}_{\mu_{b}} \frac{d\mu}{\mu}  \frac{\alpha_s(\mu)C_F}{\pi} \left [   \ln
\frac{Q^2}{\mu^2} +\ln \frac{Q^2}{P_\perp^2} \right.\nonumber\\
&& \qquad \left.
-\frac{3}{2} + c_0(R)\right ]
\ .
\end{eqnarray}
To derive the resummation result for the azimuthal angle dependent differential cross section, we first compute the Fourier transform of the soft gluon radiation contribution at one-loop order from Eq.~(\ref{int}), by applying the Jacobi-Anger expansion,
\begin{eqnarray} 
e^{iz\cos(\phi)}=J_0(z) +2 \sum_{n=1}^{\infty} i^n J_n(z) \cos(n\phi)\,,
\end{eqnarray}
and the integration formula,
\begin{eqnarray} 
\int_0^\infty \frac{d |q'_\perp|}{|q'_\perp|} J_n(|q'_\perp||b_\perp|)=\frac{1}{n}. \label{jn}
\end{eqnarray}
Importantly, the $q'_\perp$-integral gives a constant  although originally in momentum space the angular dependent terms are singular $1/q_\perp^2$, see, Eq.~(\ref{int}). At higher orders there are double logarithmic corrections but they can be resummed together with the angular-independent term   \cite{Catani:2014qha,Catani:2017tuc}.  After this resummation, we arrive at 
\begin{eqnarray}
\frac{d^5\sigma^{ep\to e'qX}}{dy_\ell d^2 P_{\perp} d^2q_{\perp}}&= &\sum_{n=1} 2 \cos (n\phi) \!\! \int \! \frac{b_\perp d b_\perp}{(2\pi)} J_n(|q_\perp|| b_\perp|) \nonumber \\
&& \times  \!\sum_q \sigma_0^{eq}  x_q  f_q(x_q,\mu_b) \frac{C_F\alpha_s c_n}{n\pi}  \nonumber \\
 &&\times e^{- \text{ Sud}^{eq}(b_\perp,P_\perp,R)}\,.  
 \label{angle} 
\end{eqnarray}
An important feature of the above result is that the Fourier coefficients  scale as 
\begin{equation}
\langle \cos (n\phi)\rangle \propto q_\perp^n\,,
\label{cosn}
\end{equation}
in the small-$q_\perp$ region~\cite{Catani:2017tuc}. 

\begin{figure}[tbp]
\begin{center}
\includegraphics[width=7cm]{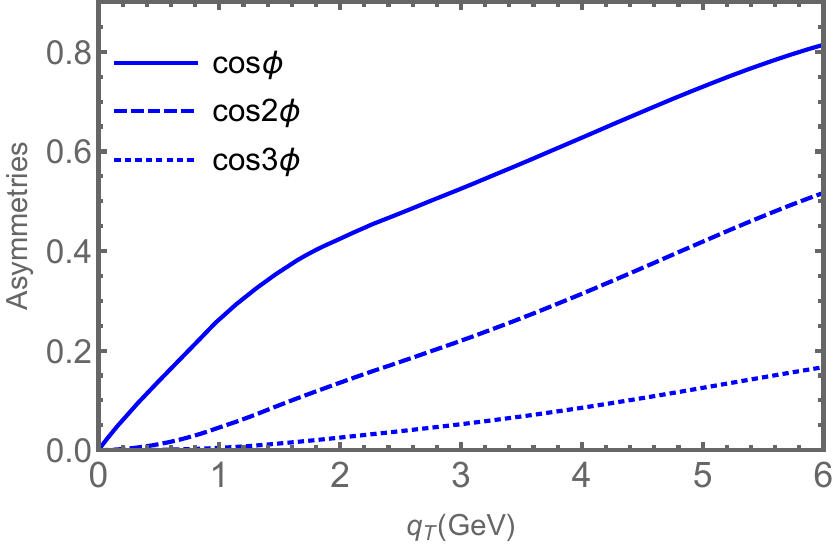}
\includegraphics[width=7cm]{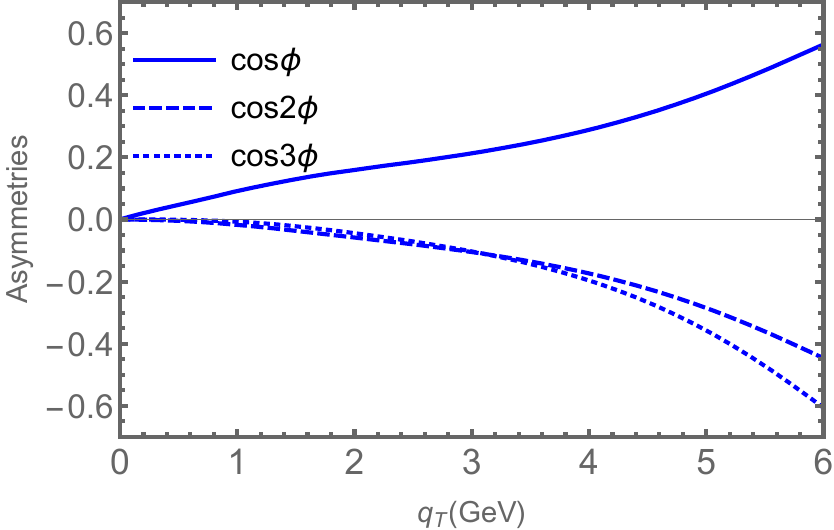}
\end{center}
\caption[*]{Azimuthal asymmetries in lepton-jet production in $ep$ collisions at $\sqrt s$=140 GeV, $P_\perp=20$ GeV, $y_l=1.5$, $Q=25$ GeV, $g_\Lambda=$0.1GeV with different jet cone sizes $R=0.4$ (top panel) and $R=1.0$ (bottom panel).}
\label{fig:ep}
\end{figure}

To evaluate (\ref{angle}), following  Ref.~\cite{Collins:1984kg} we employ the so-called $b_*$-prescription to suppress the large-$b_\perp$ region and introduce  non-perturbative form factors associated with the initial and final state radiations,
\begin{equation}
{\rm Sud}^{eq}(b_\perp)\to {\rm Sud}^{eq}(b_*)+{\rm Sud}_{\rm NP}^q(b_\perp)+{\rm Sud}_{\rm NP}^{\rm jet}(b_\perp)\ ,
\label{nonp}
\end{equation}
where $b_*=b_\perp/\sqrt{1+b_\perp^2/b_{\rm max}^2}$ with $b_{\rm max}=1.5$~GeV$^{-1}$. The form factor associated with the incoming quark is~\cite{Su:2014wpa,Prokudin:2015ysa}  
\begin{equation}
    {\rm Sud}_{\rm NP}^q(b_\perp)=0.106\, b_\perp^2+0.42\ln(Q/Q_0)\ln(b_\perp/b_*) \,,
    \label{np}
\end{equation}
with $Q_0^2=2.4$~GeV$^2$ and that for the final state jet is assumed to be
\begin{eqnarray}
\text {Sud}^{\rm jet}_{\rm NP}(b_\perp)=g_\Lambda b_\perp^2\,. \label{lambda}
\end{eqnarray}
We note that there is no  constraint for $g_\Lambda$ from experimental data so far. For an illustration, we employ the value  $g_\Lambda=0.1$ GeV$^2$. 

The numerical results for $\cos \phi$, $\cos 2\phi$ and $\cos 3\phi$ azimuthal asymmetries in typical kinematics of EIC are presented  in Fig.~\ref{fig:ep}~\footnote{For convenience, all $\cos(n\phi)$ asymmetries in Figs.~4-6 have been multiplied by a factor of $n$, i.e., the curves in these figures correspond to $\left(n\langle \cos (n\phi)\rangle\right)$.}. One can see the scaling (\ref{cosn}) in the small-$q_\perp$ region. For narrow jet ($R=0.4$, top panel) production, the $\cos \phi$ modulation is dominant as expected, though the $\cos (2\phi)$ and $\cos (3\phi)$ modulations are not negligible. Interestingly, the latter two flip signs for fat jet ($R=1$, bottom panel) production, while the $\cos \phi$ modulation is relatively unaffected. This can be understood from the numerical results of $c_n(R)$ as shown in Fig.~\ref{fig:fourier}. When the cone size $R$ increases from $0.4$ to $1$, $c_1$ remains positive, while $c_2$ and $c_3$ become negative. 

We close this subsection with a remark on the QED radiative  contribution to the  azimuthal angle asymmetries. When computing the graphs with a soft/collinear photon emitted from the final state electron, the fixed order calculation produces a large logarithm $\ln \frac{Q^2}{m_e^2}$  (roughly $\approx 19$ for typical EIC kinematics, see section V), which compensates the smallness of $\alpha_{em}$ to a large extent. This contribution can be considered as part of QED radiative corrections, similar to that discussed in Ref.~\cite{Liu:2020rvc} for inclusive DIS.

\subsection{Photon plus jet production in  $pp$ Collisions}

Next, we consider photon plus jet production in $pp$ collisions. The dominant partonic channel is $q(p_1)g(p_2)\to q(k_J)\gamma(k_\gamma)$. The leading order cross section of this process is given by,
\begin{eqnarray} 
\frac{d^6\sigma^{pp\to \gamma qX}}{d\Omega }=\sum_q \sigma_0^{qg\to\gamma q} x_q  f_q(x_q)
 x_g f_g(x_g) \delta^2(q_\perp) , \quad\quad
 \end{eqnarray}
where $\sigma_0^{qg\to\gamma q}=\frac{\alpha_s \alpha_{em} e_q^2}{N_c \hat s^2}
\left [ -\frac{\hat s}{\hat u}-\frac{\hat u}{\hat s} \right ]$ with the usual Mandelstame variables for the $2\to 2$ partonic processes: $\hat s=(p_1+p_2)^2$, $\hat t=(p_2-k_\gamma)^3$ and $\hat{u}=(p_1-k_\gamma)^2$. In the above equation $d\Omega=dy_J dy_{\gamma} d^2P_\perp
d^2 q_\perp$ represents the phase space of the final state photon and jet, and $y_\gamma$ and $y_J$ are their rapidities. The parton momenta fraction are fixed according to $x_{q,g}=P_\perp(e^{\pm y_J}+e^{\pm y_\gamma})/\sqrt{s}$. At one-loop order, the soft gluon radiation gives the following  contribution
\begin{eqnarray}
&&g^2\int \frac{d^3k_g}{(2\pi)^32E_{k_g}}\delta^{(2)} (q_\perp+k_{g\perp})
 \nonumber\\
&&~~\times\left[\frac{C_A}{2}S_g(p_1,p_2)+\frac{C_F}{2}\left(S_g(k_J,p_1)+S_g(k_J,p_2)\right)\right.\nonumber\\
&&~~ \qquad -\left.\frac{C_A-C_F}{2}\left(S_g(k_J,p_1)-S_g(k_J,p_2)\right)\right] \nonumber \\ 
&&=\frac{\alpha_s}{2\pi^2}\frac{1}{q_\perp^2}\Biggl[(C_A+C_F)\ln\frac{\hat s}{q_\perp^2}+(C_A-C_F)(y_J-y_\gamma) 
\nonumber\\
&&~~
+C_F\left(c_0+c_12\cos(\phi)+c_22\cos(2\phi) +\cdots\right)\Biggr]\  ,
\end{eqnarray}
where we used the results in Appendix B. $c_n$ are the same as in the previous subsection. Again the singularities in the azimuthally symmetric part can be resummed to all-orders in the TMD framework. Considering that the initial state consists of a quark and a gluon, we  
obtain the resummed cross section 
\begin{eqnarray} 
&&\frac{d^6\sigma^{pA\to \gamma qX}}{d\Omega}=\sum_q\int \frac{d^2 b_\perp}{(2\pi)^2} e^{i q_\perp \cdot b_\perp}\sigma_0^{qg\to \gamma q} \nonumber\\
&&\times x_q  f_q(x_q) x_g f_g(x_g) e^{- \text{ Sud}^{qg}(b_\perp,P_\perp,R) } \nonumber\\
&&\times \left ( 1+\sum_{n=1}^{\infty} \frac{\alpha_s(\mu)}{\pi} \frac{(-i)^n}{n}C_F c_n 2\cos (n\phi_b) \right )\ ,
\end{eqnarray}
where $\phi_b$ is the angle between $\vec{b}$ and $\vec{P}_\perp$. Clearly, the first term in the above contributes to the azimuthal angle averaged differential cross section. The second term contributes to various $\cos(n\phi)$ asymmetries, which can be further written as,
\begin{eqnarray} 
&&\frac{d^6\sigma^{pA\to \gamma qX}}{d\Omega}=\sum_{n=1} 2 \cos (n\phi) \!\! \int \! \frac{b_\perp d b_\perp}{(2\pi)} J_n(|q_\perp|| b_\perp|) \nonumber \\
&&~~~ \times  \!\sum_q \sigma_0^{eq}  x_q  f_q(x_q,\mu_b) x_g f_g(x_g)\frac{C_F\alpha_s c_n}{n\pi}  \nonumber \\
 &&~~~\times e^{- \text{ Sud}^{qg}(b_\perp,P_\perp,R)}\,.  
\end{eqnarray}
The perturbative Sudakov factor is given by  
\begin{eqnarray}
 &&\text{ Sud}^{qg}(b_\perp,P_\perp,R) = \int^{P_\perp}_{\mu_{b}} \frac{d\mu}{\mu} \frac{\alpha_s(\mu)}{\pi} \left \{ (C_A+C_F)  \ln
\frac{\hat s}{\mu^2}  \right.\nonumber\\
&&\left.~-2C_A  \beta_0-\!\frac{3C_F}{2}+ (C_A-C_F)(y_J-y_\gamma)\!+C_F c_0\right \},\quad
\end{eqnarray} 
where $\beta_0=11/12-N_f/18$. 

\begin{figure}[tbp]
\begin{center}
\includegraphics[width=6.7cm]{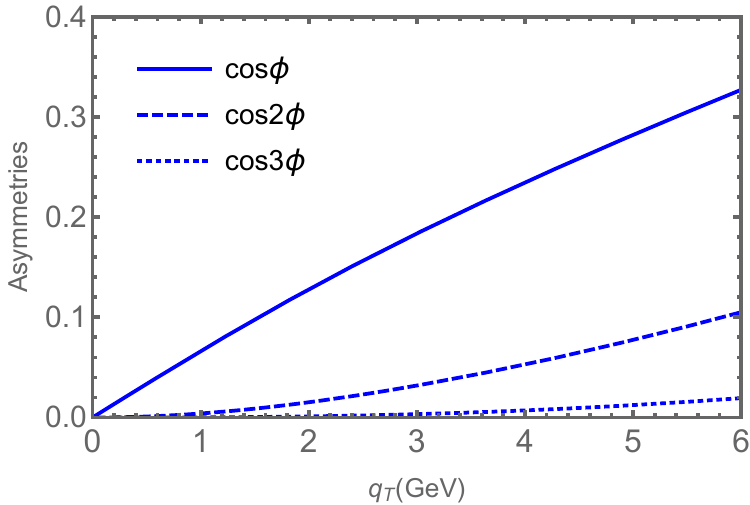}
\includegraphics[width=7cm]{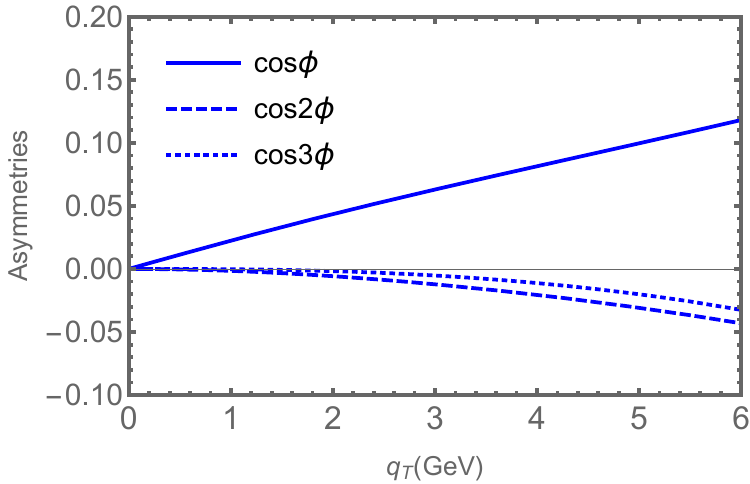}
\end{center}
\caption[*]{Azimuthal asymmetries  in photon-jet production in $pp$ collisions at RHIC as a function of $q_\perp$. $\sqrt s= 500$ GeV, $y_J=y_\gamma=1$, $P_\perp=20$ GeV, $g_\Lambda=0.1$ GeV, $R=0.4$ (top panel), $R=1$ (bottom panel)}
\label{ppRHIC}
\end{figure}
\begin{figure}[tbp]
\begin{center}
\includegraphics[width=6.8cm]{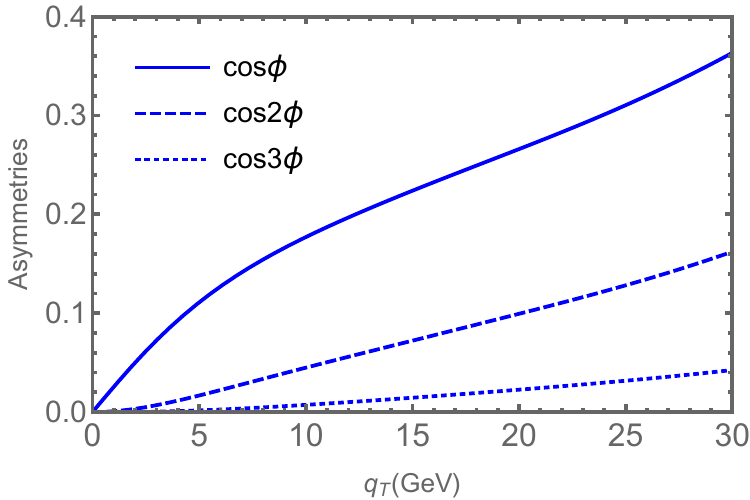}
\includegraphics[width=7cm]{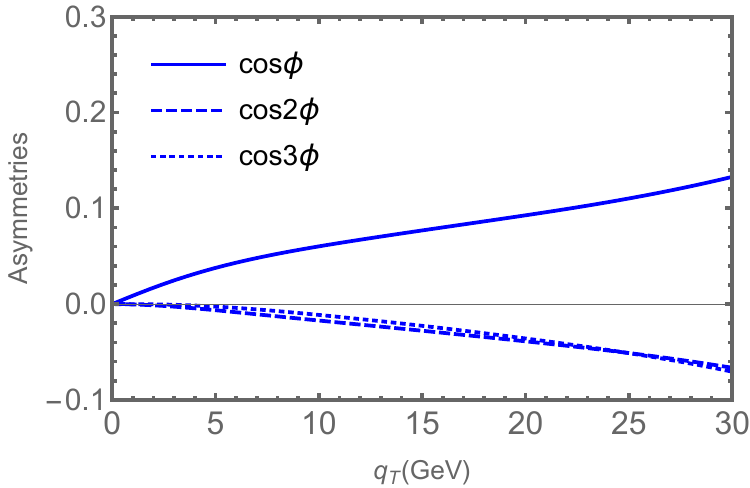}
\end{center}
\caption[*]{Azimuthal asymmetries in photon-jet production in $pp$ collisions at the LHC as a function of $q_\perp$. $\sqrt s= 7$ TeV, $y_J=y_\gamma=2$, $P_\perp=100$ GeV, $g_\Lambda=0.1$ GeV, $R=0.4$ (top panel), $R=1$ (bottom panel).}
\label{ppLHC}
\end{figure}

For the numerical evaluation, we need to introduce  non-perturbative form factors similarly to (\ref{nonp}) and (\ref{np}). In the present case, we take
\begin{equation}
{\rm Sud}^{qg}(b_\perp)\to {\rm Sud}^{qg}(b_*)+\!\frac{C_A+C_F}{C_F}{\rm Sud}_{\rm NP}^q+\!{\rm Sud}_{\rm NP}^{\rm jet}\ ,
\end{equation}  
where ${\rm Sud}_{\rm NP}^q$ is the same as (\ref{np}) with $Q\to P_\perp$. The results  for $\langle \cos (n\phi)$ ($n=1,2,3$) are shown in Fig~\ref{ppRHIC} and Fig.~\ref{ppLHC} for the RHIC and LHC kinematics, respectively.  

Similarly, for the $q\bar{q}\to g\gamma$ channel with the Born cross section
\begin{equation}
\sigma_0^{q\bar q\to \gamma g}=\frac{2\alpha_s \alpha_{em} e_q^2 C_F}{N_c \hat s^2}
\left [ \frac{\hat t}{\hat u}+\frac{\hat u}{\hat t} \right ],
\end{equation}
one can obtain the following eikonal factors due to soft gluon emissions
\begin{eqnarray}
&&~~\frac{C_A}{2} S_g(p_1,k_J)+\frac{C_F}{2}\left(S_g(p_1,p_2)+S_g(k_J,p_2)\right)\nonumber\\
&&~~ \quad -\frac{C_A-C_F}{2}\left(S_g(p_1,p_2)-S_g(k_J,p_2)\right) \nonumber \\ 
&&=2C_F\ln\frac{\hat s}{q_\perp^2}+C_A\left(c_0+\sum_{n=1}^{\infty}2c_n\cos(n\phi)\right) ,
\end{eqnarray}
which is related to the $qg\to q\gamma$ channel through the crossing symmetry ($k_J \leftrightarrow p_2$). Thus the corresponding perturbative Sudakov factor reads
\begin{equation}
\text{ Sud}^{q\bar q}= \int^{P_\perp}_{\mu_{b}} \frac{d\mu}{\mu} \frac{\alpha_s(\mu)}{\pi} \left \{ 2C_F \left[ \ln
\frac{\hat s}{\mu^2} -\frac{3}{2}\right]+C_A c_0\right \}.\,\quad
\end{equation} 
Similar numeric results can be obtained for this channel as well. However, at the RHIC and LHC kinematics, $q\bar q\to \gamma g$ channel is negilible as compared to the $qg\to \gamma q$ channel.

A few additional remarks are in order before we leave this section. First, as demonstrated in the above two cases, in which only one of the final state particles is colored while the other measured particle is color-neutral, the coefficients of odd harmonics are non-vanishing, and the dominant azimuthal angle correlation between $q_\perp$ and $P_\perp$ is of the form $\cos(\phi)$. This is simply due to the fact that the soft gluon radiation close to the measured jet is favored. The asymmetry grows with $q_\perp$ and can easily reach $10$\% or even $> 20$\% when $R$ is small. 

Second, we expect that similar conclusions should hold in other processes such as Higgs (or $Z/W$ boson) plus jet production in $pp$ collisions. For the case of the production of $Z/W$ plus jet~\cite{Sun:2018icb,Chien:2019gyf,Chien:2020hzh}, the Fourier coefficients $c_n$ and the Sudakov coefficients are identical to those in the photon plus jet case. For the case of the Higgs plus jet, the asymmetries are proportional to the same $c_n$, and the angle-independent results including the Sudakov factor can be found in Ref.~\cite{Sun:2016kkh}. 

At last, since the pure initial-state gluon radiations (represented by $S_g(p_1, p_2)$ in our calculation) do not generate asymmetries, and one of the final-state particles is colorless, one can find that only one set of the Fourier coefficients $c_n^{\text{fi}}$ (see Appendix \ref{s1} and \ref{s2}) arises from the gluon emission related to the final state jet. In contrast, a set of more complicated asymmetries depending on the rapidity difference of the final-state jets can arise from the eikonal factor $S_g(k_1, k_2)$, which involves two final-state particles. In the following section, we consider the production of two colored particles (jets) and study their angular correlations.

\section{Dijet in the Final State}

In this section we consider diffractive dijet production in $\gamma p$ or $\gamma A$ collisions and inclusive dijet production in $\gamma p$ and $pp$ collisions. Since there are two colored objects in the final state, the dominant asymmetry is expected to be the $\cos(2\phi)$ term arising from soft gluon radiation along the nearly back-to-back jets. We will only focus on $\cos(2\phi)$ asymmetry in this section. The extension to other higher harmonics (such as $\cos(4\phi)$) should be straightforward. 

\subsection{Diffractive Dijet Production}

We first study the diffractive photoproduction of dijets, $\gamma p \to q\bar{q} +p$, $\gamma A\to q\bar q+A$. In this process, an on-shell photon fluctuates into a quark-antiquark pair which then scatters off the nucleon/nucleus target via a color-singlet exchange and forms a final state dijet with momentum $k_1$ and $k_2$. This process can be studied, for example, in ultraperipheral $pA$ and $AA$ collisions at RHIC and the LHC, and the first data from the CMS collaboration came out recently \cite{CMS:2020ekd}. It can also be studied at the planned EIC in the future. Because the initial state does not carry color, there will be only final state radiation from $k_1$ and $k_2$. Therefore, the soft gluon radiation kernel is simply given by the eikonal factor
\begin{equation}
    S_g(k_1,k_2)=\frac{2k_1\cdot k_2}{k_1\cdot k_g k_2\cdot k_g}  \ .
\end{equation}
The above factor corresponds to the classical eikonal radiation from the fast-moving external currents $k_i$, which is valid in the soft limit. Integrating over the phase space (see Appendix B),  we can write 
\begin{eqnarray}
&& g^2\int \frac{d^3k_g}{(2\pi)^32E_{k_g}}\delta^{(2)} (q_\perp+k_{g\perp})
C_FS_g(k_1,k_2) \nonumber \\ 
&& =\frac{C_F\alpha_s}{\pi^2 q_\perp^2} \left [ c^{\rm diff}_0(q_\perp^2) +2 \cos (2\phi) \  c^{\rm diff}_2(q_\perp^2)+ ... \right ]. \label{feng}
\end{eqnarray}
Since the  dijet configuration is symmetric, there is no $\cos \phi$ term. Note that, for once, we have included $q_\perp$-dependence in the coefficients $c^{\rm diff}_n$ for a phenomenological reason (see below). As already mentioned in Section IIA, in general $c_n$ depends  on $q_\perp$ due to power corrections.  
Fourier  transforming (\ref{feng}) to the $b_\perp$ space and including the virtual terms, we find, at one-loop order,
\begin{eqnarray}
\tilde S^{(1)}(b_\perp)
&=&\frac{C_F\alpha_s}{\pi} c^{\rm diff}_0(0)\ln \frac{\mu_b^2}{P_\perp^2}\\
&+& \frac{C_F\alpha_s}{\pi^2}\int  d^2 q_\perp e^{iq_\perp \cdot b_\perp} \left [c^{\rm diff}_0(q_\perp^2)-c^{\rm diff}_0(0) \right ]\nonumber\\
&-&\frac{C_F\alpha_s }{\pi}2 \cos (2\phi_b) \int  d|q_\perp| 2 J_2(|b_\perp||q_\perp|)\frac{c^{\rm diff}_2(q_\perp^2)}{|q_\perp|}\ . \nonumber
\end{eqnarray} 
 After resumming the logarithms and Fourier transforming back,  we obtain  
\begin{eqnarray}
&& S(q_\perp)= \int \frac{d^2 b_\perp}{(2\pi)^2} e^{iq_\perp \cdot b_\perp} e^{-{\rm Sud}^{\rm diff.}(b_\perp,P_\perp,R)} \nonumber\\
 &&~~\times \left \{ 1+\frac{C_F\alpha_s(\mu_b)}{\pi^2} \int  d^2 q_\perp e^{iq_\perp \cdot b_\perp} \left [c^{\rm diff}_0(q_\perp^2)-c^{\rm diff}_0(0) \right ]\right \}
\nonumber \\
&&~~+\! 2 \cos (2\phi) \!\! \int \! \frac{b_\perp d b_\perp}{(2\pi)} J_2(|q_\perp|| b_\perp|) e^{-{\rm Sud}^{\rm diff.}(b_\perp,P_\perp,R)}\nonumber\\
&&\qquad ~~\times \frac{C_F\alpha_s(\mu_b)}{\pi} \! \int  d|q_\perp| 2 J_2(|b_\perp||q_\perp|)\frac{c^{\rm diff}_2(q_\perp^2)}{|q_\perp|} \ , \label{different}
\end{eqnarray}
where
\begin{equation}
    {\rm Sud}^{\rm diff.}(b_\perp,P_\perp,R)=\frac{2C_Fc^{\rm diff}_0(0)}{\pi}\int_{\mu_b}^{P_\perp} \frac{d \mu}{\mu}\alpha_s(\mu)  \ .
\end{equation}
We emphasize that the value $c^{\rm diff}_0(q_\perp=0)$ appears in the Sudakov form factor.  This is because only the leading power contribution  can be resummed into an exponential.  
To carry out the $b_\perp$-integral, we  include nonperturbative Sudakov factors. Since there is no TMD quark or gluon distribution involved, we use (see (\ref{lambda}))
 \begin{equation}
 {\rm Sud}^{\rm diff.}(b_\perp) \to {\rm Sud}^{\rm diff.}(b_*) +
2 {\rm Sud}^{\rm jet}_{\rm NP}(b_\perp)\,,
 \end{equation}  
where the factor of 2 is because there are two jets in the final state.

\begin{figure}[tbp]
\begin{center}
\includegraphics[width=7.0cm]{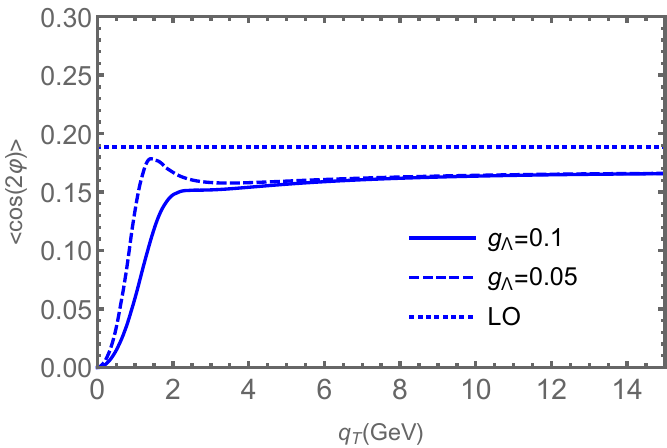}
\includegraphics[width=7.0cm]{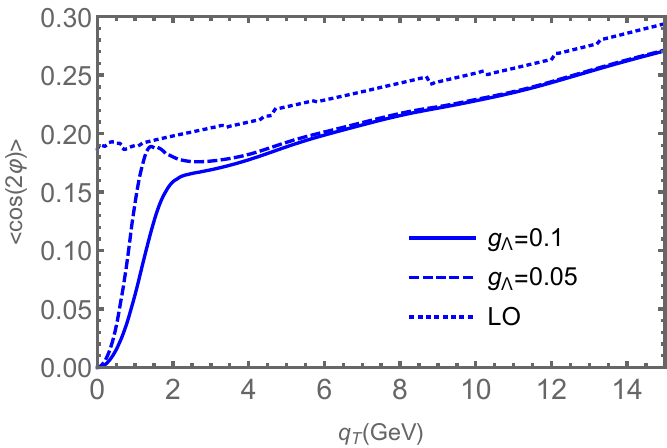}
\end{center}
\caption[*]{ Azimuthal anisotropy in diffractive dijet production  $\gamma+A\to q\bar q+A$ in ultra-peripheral heavy ion collisions at the LHC. The kinematics correspond to the CMS measurements~\cite{CMS:2020ekd} with  $P_\perp=35\, \rm GeV$, $R=0.4$ and the two jets are at the same rapidity $\Delta y_{12}=0$. The plots show $\langle \cos(2\phi)\rangle$ as a function of $q_\perp$. In the lower panel, we have included certain power corrections  $\sim q_\perp^2/P_\perp^2$ as explained in the text.  }
\label{fig:diffv2qt}
\end{figure}

Let us evaluate (\ref{different}) in two different ways. First we ignore power corrections. In this case, and in the limit $R\ll 1$, we can calculate  $c^{\rm diff}_n(q_\perp)\approx c^{\rm diff}_n(0)$ analytically 
\begin{eqnarray}
 &&c^{\rm diff}_0(0)= \ln \frac{a_0}{R^2} \ , \nonumber\\
 &&c^{\rm diff}_2(0)=  \ln \frac{a_2}{R^2} \ . \label{a0}
\end{eqnarray}
$a_0$ depends on the rapidity difference $\Delta y_{12}=|y_1-y_2|$ as $a_0=\hat{s}^2/\hat{t}\hat{u}=2+2\cosh(\Delta y_{12})$. The function $a_2(\Delta y_{12})$ remained undetermined   in our previous publication \cite{Hatta:2020bgy}, but here we can report a fully analytic result  
\begin{eqnarray}
  \ln a_2 &=& \Delta y_{12} \sinh{\Delta y_{12}} - \cosh{\Delta y_{12}} \ln\left[2\left(1+\cosh{\Delta y_{12}}\right)\right]\notag\\
  &=&-\frac{\hat u}{\hat t} \ln\frac{\hat s}{-u}-\frac{\hat t}{\hat u} \ln\frac{\hat s}{-t}\, ,\label{analytic}
\end{eqnarray}
 obtained after a rather lengthy calculation 
 outlined in Appendix B. 
 Eq.~(\ref{analytic}) has a very mild dependence on $\Delta y_{12}$. It increases slightly from $a_{2}=1/4$ to $a_{2}=1/e$ when $\Delta y_{12}$ goes from $0$ to $\infty$ \cite{Hatta:2020bgy}. 
When $R$ is not very small, we can calculate $c^{\rm diff}_n$ numerically using the formula 
\begin{eqnarray}
c^{\rm diff}_{n}(0) &=&\frac{2}{\pi}\int^{\pi/2}_{R}\frac{d \phi \, \cos 2n  \phi \,  (\pi -2 \phi)}{\sin \phi \cos \phi}  \\  
&+&\frac{2}{\pi}\int_0^{R}\frac{ d \phi\,  \cos 2n \phi}{ \sin \phi \cos \phi}\Bigl( -2 \phi +2\tan^{-1}\left[\coth y_+\tan \phi\right] \Bigr)\,, \nonumber
\end{eqnarray}
where $y_+= \sqrt{R^2-\phi^2}$ as before and we assumed $\Delta y_{12}=0$ for simplicity. In particular, 
we find for $R=0.4$, 
\begin{equation}
    c^{\rm diff}_0=3.22 \ , \qquad  \  c^{\rm diff}_2=0.60\,, \label{c02}
\end{equation}
and we may use the formula (\ref{jn}) since $c_2$ is independent of $q_\perp$.   
The leading order (LO) and resummed results for $\langle \cos(2\phi)\rangle$ are shown in the upper panel of Fig.~\ref{fig:diffv2qt}  for the LHC kinematics with $R=0.4$. The LO asymmetry is obtained from the ratio of $c^{\rm diff}_2$ to $c^{\rm diff}_0$, which is a constant in the soft gluon limit. In the small-$q_\perp$ region, we find $\langle \cos (2\phi)\rangle\propto q_\perp^2$ as expected, and in the large-$q_\perp$ region the asymmetry reaches a plateau. 

\begin{figure}[tbp]
\begin{center}
\includegraphics[width=7.0cm]{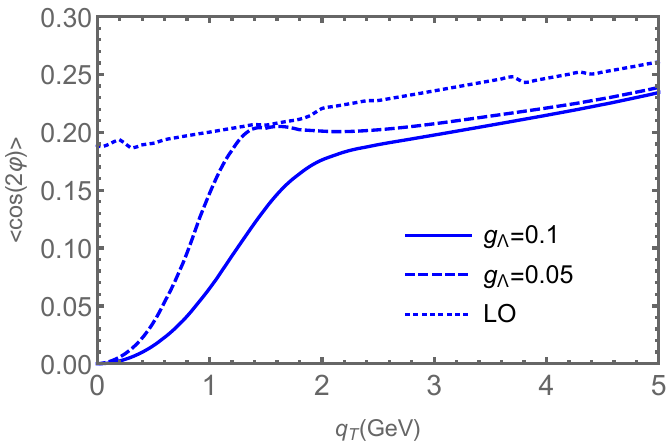}
\end{center}
\caption[*]{ Azimuthal anisotropy in diffractive dijet production  $\gamma+A\to q\bar q+A$  at the EIC with  $P_\perp=15\, \rm GeV$, $R=0.4$ and the two jets are at the same rapidity $\Delta y_{12}=0$. The plots show $\langle \cos(2\phi)\rangle$ as a function of $q_\perp$. The power corrections are included. } 
\label{fig:diffv2qtEIC}
\end{figure}

However, recent experimental data from the CMS collaboration \cite{CMS:2020ekd} show a monotonically increasing behavior in the large $q_\perp$ region. We have noticed that this discrepancy can be alleviated, at least qualitatively,  by including the following two sources of power corrections.  First, when $q_\perp$ gets larger, $|k_{1,2\perp}|$ differ from $|P_\perp|=|k_{1\perp}-k_{2\perp}|/2$ significantly.  We can  correct for this difference  by computing the soft emission kernel using the exact kinematics 
\begin{eqnarray}
S_g(k_1,k_2)
&=&\frac{2}{q_\perp^2}\frac{\cosh(y_1-y_2)-\cos(\phi_{1}-\phi_2)}{(\cosh(y_g-y_1)-\cos(\phi_1-\phi_g))}\nonumber\\
&&\times \frac{1}{(\cosh(y_g-y_2)-\cos(\phi_2-\phi_g))} \ ,
\label{powercor}
\end{eqnarray}
where $\vec{k}_{1,2\perp}=\vec{q}_\perp/2 \pm \vec{P}_\perp$, $\phi_1$, $\phi_2$ and $\phi_g$  are the  azimuthal angles of $\vec{k}_{1\perp}$, $\vec{k}_{2\perp}$ and $\vec{k}_{g\perp}$ with respect to $\vec{P}_\perp$, respectively.  Second, we impose precise rapidity cutoffs for the $y_g$-integral. For instance, when $y_1=y_2=0$, we integrate over the range $-\ln \frac{Q}{|q_\perp|}<y_g<\ln \frac{Q}{|q_\perp|}$ instead of $-\infty < y_g<\infty$. These power corrections effectively make $c^{\rm diff}_0$ and $c^{\rm diff}_2$ dependent on $q_\perp$. It is interesting to note that $c^{\rm diff}_0$ is mainly affected by the first source, whereas  $c^{\rm diff}_2$ receives contributions  from both sources with opposite signs. We have re-evaluated (\ref{different}) taking this $q_\perp$-dependence into account. The result is that  $\langle \cos (2\phi)\rangle$ now becomes an increasing function in the large-$q_\perp$ region as shown in the lower panel of Fig.~\ref{fig:diffv2qt}. The wiggles on the leading order  curve are caused by the cancellation of the two sources in $c^{\rm diff}_2$ mentioned above. Of course, this is not a fully consistent  procedure as we ignore other possible sources of power corrections such as those coming from the hard part. Yet, the better agreement with the CMS data may suggest that these are an important part  of power corrections.   In Fig.~\ref{fig:diffv2qtEIC}, we plot the asymmetry for EIC kinematics. Its size and  $q_\perp$ dependent behavior is similar to that in the CMS kinemic region except that the non-perturbative effect is more significant. 

We can also derive the resummed, full $\phi$-dependent cross section. Instead of Fourier expanding as in (\ref{feng}), we can decompose the soft factor as 
\begin{eqnarray}
 \frac{\alpha_sC_F}{(2\pi)^2}\left \{  \int \frac{dy_g \, 2 k_1 \cdot k_2}{(k_1 \cdot k_g )(k_2 \cdot k_g )}
-\frac{4c^{\rm diff}_0}{q_\perp^2} \right \}  +\frac{\alpha_sC_F}{ \pi^2}\frac{c^{\rm diff}_0}{q_\perp^2}\,,
\end{eqnarray}
where the term in the bracket is free from infrared divergence after averaging over $\phi$. The last term can be combined with the virtual contribution and get exponentiated. The resummed soft factor thus  takes form,
\begin{eqnarray}
S(q_\perp)
&= & \int \frac{d^2 b_\perp}{(2\pi)^2} e^{i b_\perp \cdot q_\perp} \left [1+\alpha_s(\mu_b) f(b_\perp) \right ] \nonumber \\ &&\times \exp\left [-\frac{\alpha_sC_F}{ \pi}\int_{\mu_b^2}^{P_\perp^2} \frac{d\mu^2}{\mu^2} c^{\rm diff}_0\right ]\,,
\label{avoid}
\end{eqnarray}
where $f(b_\perp)$ is the Fourier transform of 
\begin{eqnarray}
f(q_\perp) \!
=\! \frac{C_F}{(2\pi)^2} \! \left \{ \int \! dy_g\frac{2 k_1 \! \cdot k_2}{(k_1 \! \cdot k_g )(k_2 \cdot k_g )}
-\frac{4c^{\rm diff}_0}{q_\perp^2}\!   \right \}\,.
\end{eqnarray}
To evaluate (\ref{avoid}) efficiently,  we avoid the task of computing the  Fourier transform of $f(k_\perp)$.  Instead, we proceed by rewriting (\ref{avoid}) as 
 \begin{figure}[tbp]
\begin{center}
\includegraphics[width=7cm]{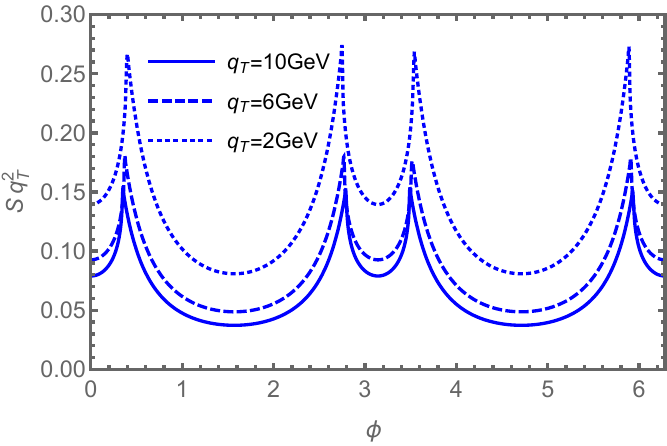}
\includegraphics[width=7cm]{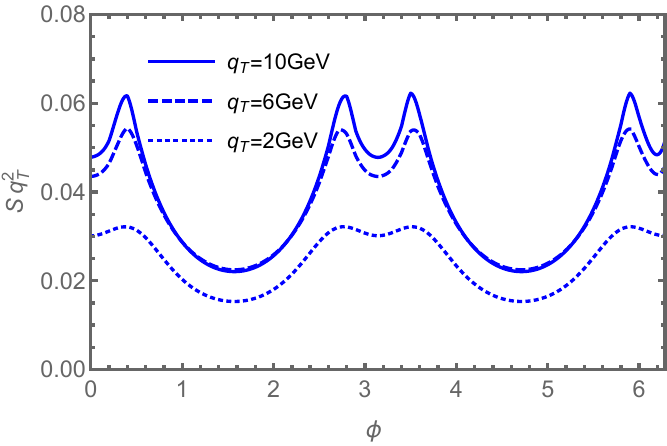}
\end{center}
\caption[*]{ Azimuthal angle distribution of $q_\perp^2 S(q_\perp)$ in diffractive dijet production  $\gamma+A\to q\bar q+A$ in ultraperipheral heavy ion collisions at the LHC. The kinematics corresponds to the CMS measurements~\cite{CMS:2020ekd} with  $P_\perp=35$  GeV, $\Delta y_{12}=0$, $R=0.4$ and $g_\Lambda=0.05$ GeV. The leading order result
(top panel) and the resummed results (bottom panel) are presented  for different values of $q_\perp$. }
\label{fig:phidis}
\end{figure}
\begin{eqnarray}
S(q_\perp)
=  S_s(q_\perp)+\int d^2 k_\perp f(k_\perp) S_a(q_\perp-k_\perp)\,, \label{kperp}
\end{eqnarray}
where
\begin{eqnarray}
&&S_s(q_\perp)=\int \frac{d^2 b_\perp}{(2\pi)^2} e^{ib_\perp \cdot q_\perp}
\exp \left [-\frac{\alpha_s}{ \pi}\int_{\mu_b^2}^{P_\perp^2} \frac{d\mu^2}{\mu^2} c^{\rm diff}_0 \right ]\nonumber\\
&&S_a(q_\perp-k_\perp)=\int \frac{d^2 b_\perp}{(2\pi)^2} e^{ib_\perp \cdot(q_\perp-k_\perp)}\alpha_s(\mu_b) \nonumber \\ && \ \ \ \ \ \ \ \ \ \ \ \ \ \ \ \ \ \ \  \times \exp \left [-\frac{\alpha_s}{ \pi}\int_{\mu_b^2}^{P_\perp^2} \frac{d\mu^2}{\mu^2} c^{\rm diff}_0 \right ]\,.
\end{eqnarray}
The subscripts $s$ and $a$ denote the azimuthally  symmetric and asymmetric parts, respectively. The $k_\perp$-integral in (\ref{kperp}) can then be done straightforwardly after including the nonperturbative Sudakov factor. 

We plot the resummed  azimuthal angle distribution of $q_\perp^2 S(q_\perp)$ in Fig.~\ref{fig:phidis} (lower panel) for different values of $q_\perp$, and also the leading order result (upper panel) for comparison . The scale of the coupling constant in the leading order calculation is chosen to be $P_\perp$. One can see that the $\phi$  distribution becomes smoother after performing the all-order resummation. 

\subsection{Inclusive Dijet in $\gamma p$ Collisions through $\gamma g\to q\bar q$}

Next we turn to inclusive dijet photoproduction $\gamma p \to jjX$. Here we focus on the direct photon contribution through $\gamma g\to q\bar q$ channel. The leading order cross section of this process is given by,
\begin{eqnarray} 
\frac{d^6\sigma^{\gamma p\to q\bar qX}}{d\Omega}=\sigma_0^{\gamma g} 
 x_g f_g(x_g) \delta^2(q_\perp) ,
 \end{eqnarray}
where $\sigma_0^{\gamma g}$ represents the leading order cross section, $d\Omega=dy_1dy_2d^2P_\perp d^2q_\perp$ for the phase space, and $f_g(x_g)$ is the gluon distribution with $x_g=P_{\perp}\left(e^{ y_1}+e^{ y_2}\right)/\sqrt{s_{\gamma p}}$ is momentum fraction of the nucleon carried by the gluon. The amplitude squared of the one soft gluon radiation can be written as~\cite{Mueller:2013wwa},
\begin{eqnarray}
{\cal M}^2&=&|\overline{\cal M}_0|^2 g_s^2\left[\frac{N_c}{2}\left(\frac{2k_1\cdot p_2}{k_1\cdot k_gp_2\cdot k_g}
+\frac{2k_2\cdot p_2}{k_2\cdot k_g p_2\cdot k_g}\right)\right.\nonumber\\
&&\left.+~~\left(-\frac{1}{2N_c}\right)\frac{2k_1\cdot k_2}{k_1\cdot k_gk_2\cdot k_g}\right] \ ,
\end{eqnarray}
where ${\cal M}_0$ represents the leading Born amplitude and the expression inside the square brackets represents the color-weighted sum of the eikonal radiation functions. All pairs external color lines need to be summed over. Applying the results from the Appendix, we obtain the soft gluon radiation contribution to the differential cross section,
\begin{eqnarray}
\left.\frac{d^6\sigma}{d\Omega}\right|_{soft}&=&\sigma_0^{\gamma g}x_gf_g(x_g)\frac{\alpha_s}{2\pi^2}\frac{1}{q_\perp^2}\biggl[C_A\ln\frac{P_\perp^2}{q_\perp^2}\nonumber\\
&&~~~+2C_F\left(c_0^{\gamma g}+c_2^{\gamma g}2\cos(2\phi)+\cdots\right)\biggr]
\ .\label{xsgp}
\end{eqnarray} 
In the small-$R$ limit, we have
\begin{eqnarray}
c_0^{\gamma g}&=&\ln\frac{a_0}{R^2} \ ,\nonumber\\
c_2^{\gamma g}&=&\frac{C_A}{2C_F}\ln\frac{a_1}{R^2}-\frac{1}{2C_FN_c}\ln\frac{a_2}{R^2} \nonumber \\ 
&=&\ln\frac{a_1}{R^2}-\frac{1}{2C_FN_c}\ln\frac{a_2}{a_1} \ , \label{radiation}
\end{eqnarray}
where $a_1=1/e$ and $a_2$ is the same as  in the previous subsection. In the following numeric calculations, we will apply $c_0^{\gamma g}=3.14$ and $c_2^{\gamma g}=0.96$ for $R=0.4$ and the two jets are at the same rapidity. 

We now perform the resummation of double logarithms in the standard TMD framework, 
\begin{eqnarray}
\frac{d^6\sigma}{d\Omega }&=&\sum_{ab}\sigma_0^{\gamma g}\int\frac{d^2\vec{b}_\perp}{(2\pi)^2}
e^{-i\vec{q}_\perp\cdot \vec{b}_\perp}\left[\widetilde{W}_{0}^{\gamma p}(|b_\perp|)\right.\nonumber\\
&&\left.~~-2\cos(2\phi_b)\widetilde{W}_{2}^{\gamma p}(|b_\perp|)\right]\ , \label{xc0}
\end{eqnarray}
where 
\begin{eqnarray}
\widetilde{W}_0^{\gamma p}(b_\perp)&=&x_g\,f_g(x_g,\mu_b)
 e^{-{\rm Sud}^{\gamma p}(P_\perp^2,b_\perp)}\ ,\\
 \widetilde{W}_2^{\gamma p}(b_\perp)&=&c_2^{\gamma g}\frac{\alpha_sC_F}{\pi}\widetilde{W}_0^{\gamma p}(b_\perp)\ . \label{wbgammap}
\end{eqnarray}
Again, we separate the Sudakov form factor into the perturbative and non-perturbative parts 
\begin{equation}
    {\rm Sud}^{\gamma p}(b_\perp,P_\perp)={\rm Sud}^{\gamma p}_{\rm pert.}(b^*,P_\perp)+{\rm Sud}^{\gamma p}_{\rm NP}(b_\perp,P_\perp) \ .
\end{equation}
The perturbative part is given by  
\begin{equation}
{\rm Sud}_{\rm pert.}^{\gamma p}=\int^{P_\perp}_{\mu_b}\frac{d\mu}{\mu}\frac{\alpha_sC_A}{\pi}
\left[\ln\frac{P_\perp^2}{\mu^2}-2\beta_0+\frac{2C_F}{C_A} c_0^{\gamma g}\right]\ , \label{su}
\end{equation}
 and we apply the $b_*$-prescription. The non-perturbative part for the present problem is 
\begin{equation}
  {\rm Sud}^{\gamma p}_{\rm NP} = \frac{C_A}{C_F}{\rm Sud}_{\rm NP}^{q} + 2{\rm Sud}_{\rm NP}^{\rm jet}\ ,
    \label{npgp}
\end{equation}
with  $Q\to P_\perp$ in (\ref{np}).  

\begin{figure}[tbp]
\begin{center}
\includegraphics[width=6.0cm]{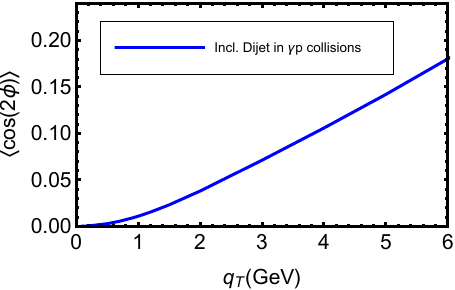}
\end{center}
\caption[*]{Anisotropy of inclusive dijet production in $\gamma p$ collisions at the future EIC for the typical kinematics: $\sqrt{S_{\gamma p}}=100 {\rm GeV}$, the leading jet $P_\perp=15~\rm GeV$ and both jets are at the same rapidity. Here we plot $\langle \cos(2\phi)\rangle$ as function and $q_\perp$, where $\phi$ is the azimuthal angle between $q_\perp$ and $P_\perp$. }
\label{fig:gpdijetv2qt}
\end{figure}

In Fig.~\ref{fig:gpdijetv2qt}, we show the numerical results for  $\langle \cos(2\phi)\rangle$ as a  function of $q_\perp$ for the typical kinematics at the future EIC, $\sqrt{s_{\gamma p}}=100\rm GeV$, $P_\perp\sim 15~\rm GeV$ and the two jets are at the same rapidity. Compared to the results in the diffractive case studied in the above subsection, the impact of resummation is more pronounced.

\subsection{Inclusive Dijet in $pp$ Collisions from $gg\to gg$ channel}

At the LHC,  dijet production is dominated by the $gg\to gg$ channel. The soft gluon radiation contribution to the azimuthally symmetric part of the differential cross section has been derived in Ref.~\cite{Sun:2015doa}. In this subsection, we  extend this work to the  angular dependent part. The leading Born amplitude can be decomposed as 
\begin{equation}
A_1f_{abe}f_{cde}+A_2f_{ace}f_{bde}+A_3f_{ade}f_{bce} \ ,
\end{equation}
where $a,b,c,d$ are color indices in the reaction $p_{1a}p_{2b}\to k_{1c}k_{2d}$. From the above color structure, we notice that $A_{1,2,3}$ represent gauge invariant amplitudes in the $s,t,u$-channels, respectively. After summing over gluon helicities, one finds the following useful results
\begin{eqnarray}
A_1^2 &=& N\frac{\hat{u}^2 +\hat{t}^2 }{\hat{s}^2}, \quad A_1 A_2^\ast =N  \frac{\hat{u}^2}{\hat{s}\hat{t}} \notag \\
A_2^2 &=& N\frac{\hat{u}^2 +\hat{s}^2 }{\hat{t}^2}, \quad A_1 A_3^\ast =-N \frac{\hat{t}^2}{\hat{s}\hat{u}} \notag \\
A_3^2 &=& N\frac{\hat{s}^2 +\hat{t}^2 }{\hat{u}^2}, \quad A_2 A_3^\ast = N \frac{\hat{s}^2}{\hat{t}\hat{u}} \, ,
\end{eqnarray}
where $N$ represents the overall normalization. The leading order amplitude squared can be written as
\begin{eqnarray}
|A_0|^2&=&\left(A_1^2+A_2^2+A_3^2+A_1A_2^*-A_1A_3^*+A_2A_3^*\right), \notag\\ 
&=&N\frac{\left(\hat{s}^2+\hat{t}^2+\hat{u}^2\right)\left(\hat{s}^4+\hat{t}^4+\hat{u}^4\right)}{2\hat{s}^2\hat{t}^2\hat{u}^2},
\end{eqnarray}
which is consistent with the well-known $gg\to gg$ Born amplitude square. The soft gluon radiation amplitude soft at one-loop order (real diagram) takes the form~\cite{Sun:2015doa},
\begin{eqnarray}
&&\frac{2k_1^\mu}{2k_1\cdot k_g}f_{gcf}\left[A_1f_{abe}f_{fde}+A_2f_{afe}f_{bde}+A_3f_{ade}f_{bfe}\right] \nonumber\\
&+&\frac{2k_2^\mu}{2k_2\cdot k_g}f_{gdf}\left[A_1f_{abe}f_{cfe}+A_2f_{ace}f_{bfe}+A_3f_{afe}f_{bce}\right] \nonumber\\
&+&\frac{2p_1^\mu}{2p_1\cdot k_g}f_{gaf}\left[A_1f_{fbe}f_{cde}+A_2f_{fce}f_{bde}+A_3f_{fde}f_{bce}\right] \ .\nonumber\\
\end{eqnarray}
Squaring this, we get 
\begin{eqnarray}
&&|A_0|^2C_A\left[S_g(p_1,p_2)+S_g(k_1,p_2)+S_g(k_2,p_2)\right]\nonumber\\
&&+\left(S_g(k_1,p_2)+S_g(k_2,p_2)-S_g(k_1,k_2)\right)\nonumber\\
&&\times \left[-\frac{N_c}{2}A_1^2-\frac{N_c}{4}\left(A_2^2+A_3^2+2A_1A_2^*-2A_1A_3^*\right)\right] \nonumber\\
&&+ \left(S_g(k_1,p_2)+S_g(p_1,p_2)-S_g(k_1,p_1)\right)\nonumber\\
&&\times\left[-\frac{N_c}{2}A_2^2-\frac{N_c}{4}\left(A_1^2+A_3^2+2A_1A_2^*+2A_2A_3^*\right)\right] \ ,\nonumber\\
&&+\left(S_g(k_2,p_2)+S_g(p_1,p_2)-S_g(k_2,p_1)\right)\nonumber\\
&&\times\left[-\frac{N_c}{2}A_3^2-\frac{N_c}{4}\left(A_1^2+A_2^2+2A_2A_3^*-2A_1A_3^*\right)\right] \ .\nonumber\\
\end{eqnarray}
By applying the following relation, 
\begin{eqnarray}
&&|A_0|^2C_A=\nonumber\\
&&=\left[-\frac{N_c}{2}A_1^2-\frac{N_c}{4}\left(A_2^2+A_3^2+2A_1A_2^*-2A_1A_3^*\right)\right] \nonumber\\
&&+ \left[-\frac{N_c}{2}A_2^2-\frac{N_c}{4}\left(A_1^2+A_3^2+2A_1A_2^*+2A_2A_3^*\right)\right] \ ,\nonumber\\
&&+\left[-\frac{N_c}{2}A_3^2-\frac{N_c}{4}\left(A_1^2+A_2^2+2A_2A_3^*-2A_1A_3^*\right)\right] \ ,
\end{eqnarray}
we can rewrite the amplitude squared  as,
\begin{eqnarray}
&&\left(S_g(p_1,p_2)+S_g(k_1,k_2)\right)\nonumber\\
&&\times \left[\frac{N_c}{2}A_1^2+\frac{N_c}{4}\left(A_2^2+A_3^2+2A_1A_2^*-2A_1A_3^*\right)\right] \nonumber\\
&&+ \left(S_g(k_2,p_2)+S_g(k_1,p_1)\right)\nonumber\\
&&\times\left[\frac{N_c}{2}A_2^2+\frac{N_c}{4}\left(A_1^2+A_3^2+2A_1A_2^*+2A_2A_3^*\right)\right] \ ,\nonumber\\
&&+\left(S_g(k_1,p_2)+S_g(k_2,p_1)\right)\nonumber\\
&&\times\left[\frac{N_c}{2}A_3^2+\frac{N_c}{4}\left(A_1^2+A_2^2+2A_2A_3^*-2A_1A_3^*\right)\right] \ .\nonumber\\
\end{eqnarray}
Furthermore, we use the following results for $A_1$, $A_2$ and $A_3$,
\begin{eqnarray}
&&\left[\frac{N_c}{2}A_1^2+\frac{N_c}{4}\left(A_2^2+A_3^2+2A_1A_2^*-2A_1A_3^*\right)\right] \nonumber\\
&&~~=\frac{N_c}{4}\frac{\hat t^2+\hat u^2}{\hat s^2-\hat t\hat u}|A_0|^2\nonumber\\
&&\left[\frac{N_c}{2}A_2^2+\frac{N_c}{4}\left(A_1^2+A_3^2+2A_1A_2^*+2A_2A_3^*\right)\right] \nonumber\\
&&~~=\frac{N_c}{4}\frac{\hat s^2+\hat u^2}{\hat s^2-\hat t\hat u}|A_0|^2\nonumber\\
&&\left[\frac{N_c}{2}A_3^2+\frac{N_c}{4}\left(A_1^2+A_2^2+2A_2A_3^*-2A_1A_3^*\right)\right] \nonumber\\
&&~~=\frac{N_c}{4}\frac{\hat s^2+\hat t^2}{\hat s^2-\hat t\hat u}|A_0|^2\ .
\end{eqnarray}
We emphasize that the above results are gauge invariant components in the amplitude squared of the $gg\to gg$ channel. Using the results of Appendix~\ref{s2}, we obtain the differential cross section 
\begin{eqnarray}
\left.\frac{d^6\sigma}{d\Omega}\right|_{soft}&=&\sigma_0^{gg}x_gf_g(x_g)\frac{\alpha_s}{2\pi^2}\frac{C_A}{q_\perp^2}\Biggl[2\ln\frac{P_\perp^2}{q_\perp^2}\nonumber\\
&&~~~+2\left(c_0^{g g}+c_2^{g g}2\cos(2\phi)+\cdots\right)\Biggr]
\ ,\label{xsgg}
\end{eqnarray} 
where $\sigma_0^{gg}$ represents the leading order cross section, $d\Omega=dy_1dy_2d^2P_\perp d^2q_\perp$ for the phase space, and $f_g(x_g)$ is the gluon distribution with $x_1=P_{\perp}\left(e^{ y_1}+e^{ y_2}\right)/\sqrt{s}$ and $x_2=P_{\perp}\left(e^{- y_1}+e^{ -y_2}\right)/\sqrt{s}$ are momentum fractions of the incoming hadrons carried by the gluons. In the small-$R$ limit, we have
\begin{eqnarray}
c_0^{gg}&=&\ln\frac{a_0}{R^2}+\frac{1}{2}\left[ \frac{\hat t^2}{\hat s^2-\hat t\hat u}\ln\frac{\hat s}{-\hat t}+\frac{\hat u^2}{\hat s^2-\hat t\hat u}\ln\frac{\hat s}{-\hat u}\right]\ ,\nonumber\\
c_2^{gg}&=&\ln\frac{a_1}{R^2}+\frac{\hat t^2+\hat u^2}{4(\hat s^2-\hat t\hat u)}\ln\frac{a_2}{a_1}\ , \label{radiationgg}
\end{eqnarray}
where $a_{0,1,2}$ are the same as in the previous sections.

The all-order resummation for dijet production in $pp$ collisions at the  next-to-leading logarithmic level has to be done in a matrix form in color space~\cite{Sun:2015doa}. This is because the final state jets and incoming partons form a color antenna with various representations of the color SU(3) group~\cite{Kidonakis:1998bk,Kidonakis:1998nf}.
For simplicity, we work in the improved leading logarithmic approximation (LLA$'$) where we include only the diagonal part in color space, namely, the leading double logarithms and those single logarithms associated with the initial parton distributions and final state jets. (The terms which depend on  kinematic variables in $c_0^{gg}$ are omitted.) In this approximation, we can write   
\begin{equation}
\widetilde{W}_0^{gg}(b_\perp)=x_1\,f_g(x_1,\mu_b)
x_2\, f_g(x_2,\mu_b) e^{-{\rm Sud}^{gg}(b_\perp,P_\perp)}\  ,\label{wb}
\end{equation}
where $f_{a,b}(x,\mu_b)$ are parton distributions for the incoming partons $a$ and $b$, and
\begin{equation}
{\rm Sud}^{gg}=\int^{P_\perp}_{\mu_b}\frac{d\mu }{\mu}
\frac{2\alpha_sC_A}{\pi}\left[\ln\left(\frac{P_\perp^2}{\mu^2}\right)-2\beta_0+\ln\frac{a_0}{R^2}\right]\ . \label{sugg}
\end{equation}
Note that only the jet-size dependent term in $c_0^{gg}$ of Eq.~(\ref{radiationgg}) was included in the above Sudakov form factor. The rest should be included in the matrix form of the resummation beyond the approximation we adopted here. For the $\cos(2\phi)$ term, we have 
\begin{eqnarray}
\widetilde{W}_2^{gg}(b_\perp)=c_2^{gg}\frac{\alpha_sC_A}{\pi}\widetilde{W}_0^{gg}(b_\perp) \ .
\end{eqnarray}
The numerical estimate of the resulting $\cos(2\phi)$ asymmetry for the LHC kinematics (after including nonperturbative Sudakov factors) has been presented in Ref.~\cite{Hatta:2020bgy}. Going beyond the LLA$'$, we need to use the matrix form of resummation. The matrix for the $\cos(2\phi)$ term may be  different from that for the azimuthally symmetric  term~\cite{Catani:2017tuc}.

For completeness, below we present the results for the other partonic channels in $pp$ collisions. Following the same procedure as shown above for the gluon channel and employing the results summarized in Appendix~\ref{s2}, one can obtain the corresponding expression for the $qq^\prime \to qq^\prime$ channel as follows
\begin{eqnarray}
\left.\frac{d^6\sigma}{d\Omega}\right|_{soft}&=&\sigma_0^{qq}x_qf_q(x_q)\frac{\alpha_s}{2\pi^2}\frac{C_F}{q_\perp^2}\Biggl[2\ln\frac{P_\perp^2}{q_\perp^2}\nonumber\\
&&~~~+2\left(c_0^{qq}+c_2^{qq}2\cos(2\phi)+\cdots\right)\Biggr]
\ .\label{xsqq0}
\end{eqnarray} 
In the small-$R$ limit, one finds
\begin{eqnarray}
c_0^{qq}&=&\ln\frac{a_0}{R^2}+\left[ \frac{N_c^2+1}{N_c^2-1}\ln\frac{\hat s}{-\hat t}-\frac{N_c^2-3}{N_c^2-1}\ln\frac{\hat s}{-\hat u}\right]\ ,\nonumber\\
c_2^{qq}&=&c_2^{\text{fi}}+\frac{1}{N_c C_F}\left(c_2^{\text{ff}}-c_2^{\text{fi}}\right)=\ln\frac{a_1}{R^2}+\frac{1}{4}\ln\frac{a_2}{a_1}\ , \label{radiationqq}
\end{eqnarray}
where $c_0^{qq}$ agrees with the result in Ref.~\cite{Sun:2015doa} with $N_c=3$.

Similarly, the results for the $gg\to q\bar{q}$ can be cast into
\begin{eqnarray}
\left.\frac{d^6\sigma}{d\Omega}\right|_{soft}&=&\sigma_0^{q\bar{q}}x_gf_g(x_g)\frac{\alpha_s}{2\pi^2 q_\perp^2}\Biggl[2C_A \ln\frac{P_\perp^2}{q_\perp^2}\nonumber\\
&&~~~+2C_F\left(c_0^{q\bar{q}}+c_2^{q\bar{q}}2\cos(2\phi)+\cdots\right)\Biggr]
\ .\label{xsqq}
\end{eqnarray} 
In the small-$R$ limit, one finds the expressions for the coefficients
\begin{eqnarray}
c_0^{q\bar{q}}&=&\ln\frac{a_0}{R^2}+\frac{N_c^2}{2C_F}\left[ \frac{ \hat{t}^2\ln\frac{\hat s}{-\hat t} +\hat{u}^2\ln\frac{\hat s}{-\hat u} }{C_F \hat{s}^2- N_c \hat{t}\hat{u}}\right] \\
c_2^{q\bar{q}}&=&\ln\frac{a_1}{R^2}+\left[\frac{\hat{s}^2+2N_c^2\hat{u}\hat{t}}{4N_c^2C_F(C_F \hat{s}^2- N_c \hat{t}\hat{u})} \right]\ln\frac{a_2}{a_1}\ . \label{radiationqqbar}
\end{eqnarray}
For the inverse process of the above channel, i.e., $q\bar q \to gg$, one simply can replace the color factors of $\ln\frac{P_\perp^2}{q_\perp^2}$ and Fourier coefficients by $C_F$ and $C_A$ in Eq.~(\ref{xsqq}). Then the first two coefficients are 
\begin{eqnarray}
c_0^{q\bar{q} \to gg}&=&\ln\frac{a_0}{R^2}+ \frac{C_F-N_c}{N_c}\ln \frac{\hat{s}^2}{\hat t \hat u}+\frac{N_c}{2}\frac{ \hat{t}^2\ln\frac{\hat s}{-\hat t} +\hat{u}^2\ln\frac{\hat s}{-\hat u}  }{C_F \hat{s}^2- N_c \hat{t}\hat{u}} \notag \\ 
c_2^{q\bar{q}\to gg}
&=&\ln\frac{a_1}{R^2}+\left[\frac{N_c\left(\hat{u}^2+\hat{t}^2\right)}{4(C_F \hat{s}^2- N_c \hat{t}\hat{u})} \right]\ln\frac{a_2}{a_1}. 
\end{eqnarray}

For the $qg\to qg$ channel, we have the following soft gluon radiation contribution to the differential cross section,
\begin{eqnarray}
\left.\frac{d^6\sigma}{d\Omega}\right|_{soft}&=&\sigma_0^{qg}x_qf_q(x_q)x_gf_q(x_g)\frac{\alpha_s}{2\pi^2}\frac{C_F+C_A}{q_\perp^2}\label{xsqg}\\
&&\times \Biggl[\ln\frac{P_\perp^2}{q_\perp^2}+c_0^{qg}+c_2^{qg}2\cos(2\phi)+\cdots\Biggr]
\ .\nonumber
\end{eqnarray} 
Taking the small-$R$ limit, one finds
\begin{eqnarray}
c_0^{qg}&=&\ln\frac{a_0}{R^2}+ \frac{ \left(-20 \hat{s}^2+5 \hat{s} \hat{u}+61 \hat{u}^2\right)}{13 \left(4 \hat{s}^2-\hat{s} \hat{u}+4 \hat{u}^2\right)}\ln \frac{\hat{s}}{-\hat{u}}\nonumber\\
&&+ \frac{ \left(11 \hat{s}^2-23 \hat{s}\hat{u}+11 \hat{u}^2\right)}{13 \left(4 \hat{s}^2-\hat{s} \hat{u}+4 \hat{u}^2\right)} \ln\frac{\hat{s}}{-\hat{t}}\ ,\nonumber\\
c_2^{qg}&=&\ln\frac{a_1}{R^2}+\frac{9 (9\hat{u}^2-\hat{t}^2) }{26 \left(4 \hat{s}^2-\hat{s} \hat{u}+4 \hat{u}^2\right)} \ln\frac{a_2}{a_1}\ , \label{radiationqg}
\end{eqnarray}
where $c_0^{qg}$ also agrees with the result in Ref.~\cite{Sun:2015doa} with $N_c=3$ and $c_2^{qg}$ is new. Here we have neglected the odd harmonics since those terms vanish when final state jets are symmetrized. 

Let us comment on the patterns of the above logarithms that one observes from the various processes. First, the color factors of the term $\ln\frac{P_\perp^2}{q_\perp^2}$ are associated with the incoming partons, while the color factor of the logarithm $\ln\frac{a_{0, 1}}{R^2}$ are determined by the final state jets. The second term of the coefficients $c_0$ and $c_2$ are process dependent. In particular, the  $\ln\frac{a_2}{a_1} $ terms are expected to be small and it vanishes when $\Delta y_{12} \to \infty$.

\section{Dijet Production in DIS to Probe the linearly polarized gluon distribution}

In this section, we return to 
 inclusive dijet production in DIS off a nucleon/nucleus
\begin{equation}
    e+A(p_A)\to e'+jet_1 (k_1)+jet_2 (k_2) +X \ . 
\end{equation}
Differently from Section IIIB, here the exchanged photon is virtual with invariant mass squared $q^2=-Q^2$.   
As mentioned in the Introduction, this process has been proposed to study one particular aspect of the gluon distribution in the nucleon/nucleus,  the so-called linearly polarized gluon distribution~\cite{Boer:2010zf,Metz:2011wb,Dominguez:2011br,Dumitru:2015gaa,Mantysaari:2019hkq}.  The dependence of the differential cross section on $q_\perp=k_{1\perp}+k_{2\perp}$ is sensitive to the TMD gluon distributions of the nucleon/nucleus. Among them, the linearly polarized gluon distribution will lead to a characteristic $\cos(2\phi)$ asymmetry, where $\phi$ is the azimuthal angle between $\vec{q}_\perp$ and $\vec{P}_\perp$~\cite{Boer:2010zf}. This observation has gained more importance  when it was realized that the linearly polarized gluon distribution is of the same size as the usual gluon TMD distribution in the small-$x$ saturation formalism~\cite{Metz:2011wb}. Since then, several proposals have been made to measure the $\cos (2\phi)$ asymmetry at the planned electron-ion collider~\cite{Dumitru:2015gaa,Mantysaari:2019hkq}. Moreover, the  distribution has been widely applied to many other processes~\cite{Qiu:2011ai,Boer:2011kf,Akcakaya:2012si,Pisano:2013cya,Boer:2013fca,Boer:2014lka,Boer:2016fqd,Boer:2017xpy,Boer:2020bbd}. 

However, there are two important issues which complicate the interpretation of such measurements, but have not been adequately investigated in the literature. First, the collinear gluon radiation from the incoming parton can generate the  $\cos(2\phi)$ modulation. This  has been well understood in the collinear factorization framework at moderate transverse momentum and also in the TMD resummation formalism~\cite{Collins:1984kg,Nadolsky:2007ba,Catani:2010pd,Sun:2011iw,Gutierrez-Reyes:2019rug}. This perturbative effect can mimic the nonperturbative, intrinsic modulation due to the linearly polarized gluon distribution, but it has been largely ignored in the previous phenomenological  studies. Another source of the $\cos (2\phi)$ correlation that has been missing in the literature of the linearly polarized gluon distribution is the soft gluon radiation from the final state jets as pointed out in \cite{Hatta:2020bgy} and discussed in the previous sections. Therefore, in order to reliably extract the linearly polarized gluon distribution through the measurement of the $\cos (2\phi)$ asymmetry,  it is important to quantify these `background' effects.\footnote{Recent studies \cite{Altinoluk:2021ygv,Boussarie:2021lkb} show that power-corrections in the hard part can also  affect the $\cos (2\phi)$ asymmetry from the linearly polarized gluon distribution . }

In this section, we perform a systematic study of the $\cos(2\phi)$ asymmetry in DIS dijet production  including the above three physics: the  `intrinsic' and `collinear radiation generated' linearly polarized gluon distributions, and the final state soft gluon radiation contribution. We shall focus on the TMD domain, i.e., $P_\perp\gg q_\perp$, where the leading jet transverse momentum is much larger than the total transverse momentum of the two jets. In this region, we have to perform an all-order resummation of the logarithms $(\alpha_s \ln^2 P^2_\perp/q_\perp^2)^n$. 

Previously, the TMD resummation for the linearly polarized gluon distribution has been studied in Refs.~\cite{Nadolsky:2007ba,Catani:2010pd,Sun:2011iw} and its impact on  Higgs Boson production was found to be very small~\cite{Wang:2012xs,Boer:2014tka}. Resummation effects on the $\cos(2\phi)$ in photon-jet correlation have also been studied in Ref.~\cite{Boer:2017xpy}, which, however, only included the intrinsic linearly polarized gluon contribution.  Our calculations in the following will show that the collinear radiation generated linearly polarized gluon distribution dominates over the intrinsic one  at higher hard momentum scales $Q^2$, say, at the scale of the Higgs mass. However, their relative importance strongly depends on $Q^2$, and this leaves an opportunity to explore the transition from intrinsic to collinear radiation regimes in future experiments.

\subsection{Linearly Polarized Gluon Distribution}

In this subsection, we briefly introduce the linearly polarized gluon distribution and study the associated QCD evolution. The TMD gluon distributions are defined through the following matrix element~\cite{Collins:1981uw,Mulders:2000sh,Ji:2005nu},
\begin{eqnarray}
&&{\cal M}^{\mu\nu}(x,k_\perp)=\int\frac{d\xi^-d^2\xi_\perp}{P^+(2\pi)^3}
    e^{-ixP^+\xi^-+i\vec{k}_\perp\cdot \vec\xi_\perp}\label{kg1}\\
    &&\times 
    \langle P|{F_a^{+\mu}}(\xi^-,\xi_\perp)
{\cal L}^\dagger_{vab}(\xi^-,\xi_\perp) {\cal L}_{vbc}(0,0_\perp)
F_c^{\nu+}(0)|P \rangle\ ,\nonumber
\end{eqnarray}
where the nucleon moves along $+\hat z$-direction and $F^{\mu\nu}_a$ is the gluon field strength tensor. The light-cone components are defined as $k^\pm=(k^0\pm k^3)/\sqrt{2}$. In the above equation, $x$ is the longitudinal momentum fraction carried by the gluon and $k_\perp$ is the transverse momentum. The gauge link ${\cal L}_v$ is constructed in the adjoint representation and depends on the process~\cite{Dominguez:2010xd}.  For an unpolarized nucleon at leading twist, the above matrix element contains two independent TMD gluon distributions
~\cite{Mulders:2000sh},
\begin{eqnarray}
{\cal M}^{\mu\nu}(x,k_\perp)&=&\frac{1}{2}\Biggl[xf_g(x,k_\perp) g_\perp^{\mu\nu}\nonumber\\
&&+xh_g(x,k_\perp)\left(\frac{2k_\perp^\mu k_\perp^\nu}{k_\perp^2}-g_\perp^{\mu\nu}\right) \Biggr]\ ,\label{kg2}
\end{eqnarray}
where $g_\perp^{\mu\nu}$ has only transverse components $g_\perp^{ij}=\delta^{ij}$. $f_g(x,k_\perp)$ is the usual azimuthally symmetric TMD gluon distribution, and $h_g(x,k_\perp)$ is the linearly polarized gluon distribution. $h_g$ vanishes when ${\cal M}^{\mu\nu}$ is integrated  over transverse momentum, which means there is no integrated version of the linearly polarized gluon distribution $h_g(x)$. 

To apply  TMD gluon distributions in hard scattering processes, we have to take into account the TMD evolution and  resummation. For the azimuthally symmetric part, we follow the ``standard" scheme (also called Collins 2011 Scheme)~\cite{Collins:2011zzd,Catani:2000vq,Catani:2013tia,Prokudin:2015ysa} which reads, in coordinate space, 
\begin{eqnarray}
\widetilde{f}_g(x,b_\perp,\zeta_c=Q^2)&=& e^{-{\rm Sud}^g_{\rm pert}(Q^2,b^*)-{\rm Sud}^g_{\rm NP}(Q,b_\perp)}  \label{56}\\
&&\times {\widetilde{\cal F}}_g(\alpha_s(Q))\sum_{i}C_{g/i}\otimes f_i(x,\mu_b) \,, \nonumber
\end{eqnarray}
where the perturbative Sudakov factor is process-dependent (to be specified below) and the nonperturbative part is given by ${\rm Sud}^g_{\rm NP}=\frac{C_A}{C_F}{\rm Sud}^q_{\rm NP}$. We have set the rapidity regulator $\zeta_c$ and the renormalization scale $\mu^2$ to be both $Q^2$, and ${\widetilde{\cal F}}_g(\alpha_s(Q))=1+{\cal O}(\alpha_s)$ in the standard TMD scheme. For the $C$-coefficients, we use the one-loop results 
\begin{eqnarray}
C_{g/g}&=&\delta(1-x)+ {\cal O}(\alpha_s^2) \label{cgg},\\
C_{g/q}&=& \frac{\alpha_sC_F}{2\pi}x \ ,
\end{eqnarray}
in the standard TMD scheme. Note that in this scheme $C_{g/g}$ vanishes at one-loop order. Numerically, we find that the contribution from $C_{g/q}^{(1)}$ is negligible. We therefore only keep the delta function term in (\ref{cgg}) in the following. 

The Collins-Soper evolution equation for the linearly polarized gluon distribution can be derived in a similar manner.  Again in the $b_\perp$-space, we parametrize as  
\begin{eqnarray}
\tilde h_g^{\mu\nu}(x,b_\perp)&=&\frac{1}{2}\left(g_\perp^{\mu\nu}-\frac{2b_\perp^\mu b_\perp^\nu}{b_\perp^2}\right)\tilde{h}_g(x,b_\perp) \ , 
\end{eqnarray}
where the tensor structure is uniquely determined by the traceless condition.  The solution of the evolution equation takes the form
\begin{equation}
\widetilde{h}_g(x,b_\perp;\zeta_c=Q^2)= e^{-{\rm Sud}_{\rm pert}^g(Q^2,b_*)}\widetilde{h}_g(x,b_\perp,\zeta_c=\mu_b) \ .
\label{resum}
\end{equation}
The distribution at the lower scale $\zeta_c=\mu_b$ does not contain large logarithms, but it cannot be written similarly to  (\ref{56}) because  there is no integrated $h_g$  distribution. 
In the large-$b_\perp$ region where physics becomes nonperturbative, it has to be modeled. Following Ref.~\cite{Boer:2011kf}, we parameterize the linearly polarized gluon distribution in term of the normal gluon distribution, 
\begin{eqnarray}
    \widetilde{h}_g(x,b_\perp,\mu_b)|_{b_\perp\gg \Lambda^{-1}_{QCD}}&=& 
    \frac{eQ_h^2b_\perp^2}{27} e^{\frac{b_\perp^2Q_h^2}{12}}e^{-{\rm Sud}_{\rm NP}^g(Q,b_\perp)}\nonumber\\
    &&\times f_g(x,\mu_b) \ ,\label{eq:e9}
\end{eqnarray}
where $Q_h\approx 1~\rm GeV$ and $f_g(x,\mu)$ is the integrated gluon distribution. In this model, we assume that the linearly polarized gluon distribution has the same $x$-dependence as the normal gluon distribution $f_g(x,\mu)$. In reality, they may be totally different. Eq.~(\ref{resum}) with (\ref{eq:e9}) (extrapolated to the full $b_\perp$ region including small-$b_\perp$) is what we call the intrinsic part of the linearly polarized gluon distribution. In momentum space, it is proportional to $k_\perp^2$ in the small $k_\perp$ region and satisfies the positivity bound  $h_g<f_g$.  As mentioned already,  in most literature only this part has been   used to calculate  the $\cos (2\phi)$ asymmetry. In such approaches, $\tilde{h}_g(b_\perp)\sim b_\perp^2$ as $b_\perp\to 0$.  

However, we know that the small-$b_\perp$ behavior of the linearly polarized gluon distribution is  perturbatively calculable via collinear gluon radiation at large transverse momentum 
\begin{eqnarray}
    &&\widetilde{h}_g(x,b_\perp,\zeta_c=\mu_b)|_{\mu_b\gg\Lambda_{QCD}}=\sum_i\int\frac{dx'}{x'}f_{i}(x',\mu_b)\nonumber\\
    &&~~~~~~~~~~~~~~~~~~~~~\times C_{h/i}(x/x',\mu_b)  \ .\label{eq:e10}
\end{eqnarray}
The $C$-coefficients start at ${\cal O}(\alpha_s)$, and are given by \cite{Nadolsky:2007ba,Catani:2010pd,Sun:2011iw,Gutierrez-Reyes:2019rug}
\begin{eqnarray}
    C_{h/q}&=&\frac{\alpha_s}{2\pi}C_F\frac{1-\xi}{\xi}+{\cal  O}(\alpha_s^2) \ , \\
    C_{h/g}&=&\frac{\alpha_s}{2\pi}C_A\frac{1-\xi}{\xi}+{\cal O}(\alpha_s^2) \ . 
\end{eqnarray}
The two-loop results have been recently derived in Ref.~\cite{Gutierrez-Reyes:2019rug}.  In the following, as an illustration we  only use  the one-loop results.   
We see from (\ref{eq:e10}) that the correct small-$b_\perp$ behavior is not quadratic but constant (up to the logarithmic running of the coupling)  
\begin{equation}
    \widetilde{h}_g(x,b_\perp)_{b_\perp\ll 1/\Lambda_{QCD}}\propto \alpha_s(\mu_b) \ . 
\end{equation}
This is so because there is no dimensionful parameter in perturbation theory to compensate the dimension of $b_\perp^2$. 

Comparing the large and small $b_\perp$ behaviors of the intrinsic (\ref{eq:e9}) and radiative  (\ref{eq:e10}) contributions, we notice that they dominate $\widetilde{h}_g(x,b_\perp)$ separately in these two regions. Therefore, we can combine them together and arrive at the following two-component model 
\begin{eqnarray}
&&\widetilde{h}_g(x,b_\perp;\zeta_c=Q^2)= e^{-{\rm Sud}_{\rm pert}^g(Q^2,b^*)-{\rm Sud}_{\rm NP}^g(Q,b_\perp)} \label{twocomp}\\
&&~~~\times \left[g_h(b_\perp) f_g(x,\mu_b)+\int\frac{dx'}{x'}f_{i}(x',\mu_b) C_{h/i}(x/x',\mu_b) \right]\nonumber\ . 
\end{eqnarray}
At large $b_\perp$, the first term dominates, while at small-$b_\perp$, the second term dominates and the first term is negligible.

\begin{figure}[tbp]
\begin{center}
\includegraphics[width=6cm]{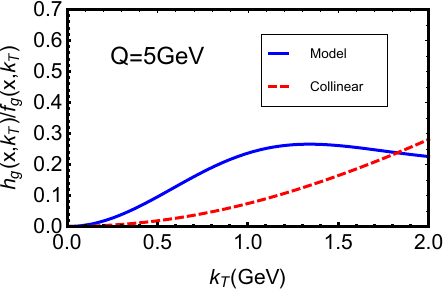}
\includegraphics[width=6cm]{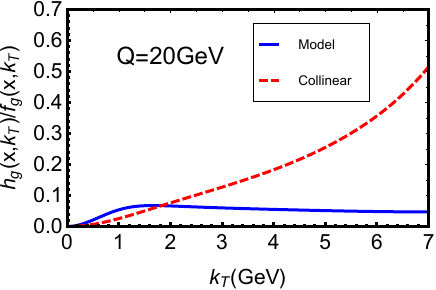}
\includegraphics[width=6cm]{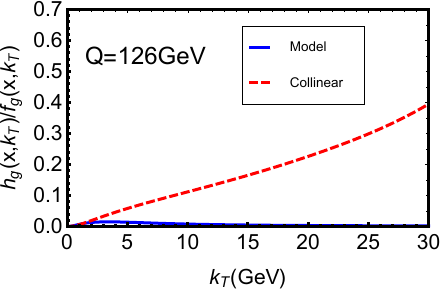}
\end{center}
\caption[*]{The ratio of the linearly polarized gluon distribution over the symmetric one ${h}_g(x,k_\perp,\zeta_c=Q)/f_g(x,k_\perp,\zeta_c=Q)$ as a functions of $k_\perp$ for different values $Q=5$, $20$, $126~\rm GeV$ with $x=0.05$. The blue and red curves are the intrinsic and radiative parts, respectively. }
\label{fig:hgqkt}
\end{figure}

An important feature of (\ref{twocomp}) is that the scale ($Q^2$) dependence of the linearly polarized gluon distribution is dictated by the Sudakov form factor. At very large $Q^2$, it pushes the $b_\perp$-distribution to the small-$b_\perp$ region, and we only need to take into account the contribution from the second term. At low $Q^2$, the large-$b_\perp$ region will be important to obtain the $k_\perp$ distribution, and one may keep only the first term to a good approximation. For  moderate values of $Q^2$, we may need to take into account both contributions. Therefore, by studying different hard processes sensitive to the linearly polarized gluon distribution, we will be able to investigate the transition from non-perturbative  to perturbative regimes. This provides a unique perspective for the nucleon/nucleus tomography study in future experiments, in particular,  at the planned EIC.

In Fig.~\ref{fig:hgqkt}, we show the ratio $h_g(x,k_\perp)/f_g(x,k_\perp)$ as a function of $k_\perp$ for different  values of $Q^2$ at a fixed value of $x=0.05$. These plots clearly demonstrate the above points.  In particular, it is interesting to notice that, for Higgs boson production at the scale $Q=M_h$, the linearly polarized distribution is completely dominated by the collinear radiation contribution. On the other hand, when $Q=5$ GeV, the $k_\perp$-dependence is dominated by the non-perturbative part. Of course, we do not know the actual magnitude of the nonperturbative part. (Eq.~(\ref{eq:e9}) is just a model.) However, our results  suggest that this can be constrained by scanning $Q^2$ in the relatively low momentum region at the  EIC. 

In the above discussions, the separation of the ``intrinsic" and ``collinear" parts of the linearly polarized gluon distribution depends on the model we used, where they have different behaviors at large and small $b_\perp$. It will be interesting to develop a model to capture both features in a single set-up. In particular, in the small-$x$ dipole formalism, both the linearly polarized gluon distribution and the normal gluon distribution can be calculated from the same dipole amplitude~\cite{Metz:2011wb} and it may be possible to include the ``intrinsic" and ``collinear" parts at the same time. Further developments are needed to implement collinear gluon radiation contribution in the gluon distribution functions in the small-$x$ dipole formalism, see, for example, the discussions in Ref.~\cite{Xiao:2017yya}.

\subsection{$\cos 2\phi$ Correlation in Inclusive Dijet Production in DIS}

We now calculate the $\cos(2\phi)$ asymmetry in inclusive dijet production in DIS  using the linearly polarized gluon distribution constructed in the previous subsection.  
We focus on the gluonic channel  $\gamma^*_{T,L} g\to q\bar q$ where the incoming photon can be transversely ($T$) or longitudinally ($L$) polarized. In the collinear factorization framework, the differential cross section can be written as,
\begin{equation}
    \frac{d^6\sigma^{T,L}}{d\Omega}=\int\frac{dx'}{x'}x_gf_g(x')\left[\hat\sigma_{0}^{T,L}+2\cos(2\phi)\hat\sigma_{2}^{T,L}\right] \ , \label{crosssection}
\end{equation}
where 
$d\Omega=dy_1dy_2d^2P_\perp d^2q_\perp$. $f_g$ represents the integrated gluon distribution. At the leading Born order, we have 
\begin{equation}
    \hat\sigma_{0}^{T,L}=\sigma_{0}^{T,L}\delta^{(2)}(q_\perp)\delta(\xi-1) \ ,~\hat\sigma_{2}^{T,L}=0
\end{equation}
where $\xi=x_g/x'$ with $x_g$ being the momentum fraction  carried by the gluon. It  can be determined from the dijet kinematics as $x_g=(\frac{P_\perp^2}{z(1-z) }+Q^2)/(s+Q^2)$ where $s$ is the center of mass energy of the $\gamma^* p $ system and $z$ is the momentum fraction of the virtual photon carried by the quark  jet. 

Gluon radiations from the incoming gluon and the outgoing $q\bar{q}$ pair  will generate not only a nonzero transverse momentum $q_\perp$ but also a $\cos(2\phi)$ asymmetry. In the TMD kinematics $P_\perp \gg q_\perp$, we have both collinear and soft gluon contributions,
\begin{eqnarray}
\hat\sigma_0^{(1)}&=&\sigma_0\frac{\alpha_s}{2\pi^2}\frac{1}{q_\perp^2}\left[{\cal P}_{g/g}^{(<)}(\xi)+\delta(1-\xi)\left(C_A\ln\frac{P_\perp^2}{q_\perp^2}\right.\right.\nonumber\\
&&\left.\left.~+2C_Fc_0^{\gamma g} 
+2C_A\ln\frac{\hat{s}+Q^2}{\hat s}\right)\right]  \label{eq71} \ ,\\
\hat\sigma_2^{(1)}&=&\sigma_2\frac{\alpha_s}{2\pi^2}\frac{1}{q_\perp^2}\left[C_A\frac{1-\xi}{\xi}+\frac{\sigma_0}{\sigma_2}\delta(1-\xi)2C_Fc_2^{\gamma g} \right] \ , 
\end{eqnarray}
where $\sigma_0$ and $\sigma_2$ are normalization factors for the differential cross sections and $c_0^{\gamma g}$ and $c_2^{\gamma g}$ are the same as those defined for $\gamma g\to q\bar q $ subprocess in Sec.~IIIB. ${\cal P}_{g/g}^{(<)}$ denotes the collinear splitting kernel without the delta function part. The soft radiation part is essentially the same as (\ref{radiation}) except that now the phase space has increased $\hat{s}\to \hat{s}+Q^2$ since the incoming photon is virtual. Note that the Mandelstam variables are related by $\hat s \hat t \hat u = P_\perp^2 (\hat s +Q^2)^2$ in this case. The above equations apply to both transverse and longitudinal incoming photons. To leading order, we have the  following relations for the cross section ratios~\cite{Metz:2011wb,Dominguez:2011br},
\begin{equation}
    \frac{\sigma_2^T}{\sigma_0^T}=- \frac{\epsilon_f^2P_\perp^2}{\epsilon_f^4+P_\perp^4}\ ,\qquad \frac{\sigma_2^L}{\sigma_0^L}=\frac{1}{2} \ ,\label{ratio}
\end{equation}
where $\epsilon_f^2=z(1-z)Q^2$.

The singularity $q_\perp \to 0$ can be factorized and resummed in the $b_\perp$-space into the TMD gluon distributions  $f_g$ and $h_g$, as well as the soft factors associated with the final state jets. This converts (\ref{crosssection}) into 
\begin{eqnarray}
\frac{d^6\sigma}{d\Omega }&=&
\sigma_0\int\frac{d^2\vec{b}_\perp}{(2\pi)^2}
e^{-i\vec{q}_\perp\cdot \vec{b}_\perp}\left[\widetilde{W}_{0}^{\gamma^* p}(|b_\perp|)\right.\nonumber\\
&&\left.~~-2\cos(2\phi_b)\widetilde{W}_{2}^{\gamma^* p}(|b_\perp|)\right]\ , \label{xc1}
\end{eqnarray}
where the azimuthal symmetric term can be written as
\begin{equation}
\widetilde{W}_0^{\gamma^* p}(b_\perp)=x_g\,f_g(x_g,\mu_b)
 e^{-{\rm Sud}_{\rm pert}^{\gamma^* p}(b_*)-{\rm Sud}_{\rm NP}^{\gamma^*p}(b_\perp)}\  .\label{wb0}
\end{equation}
The perturbative Sudakov form factor is defined as
\begin{eqnarray}
{\rm Sud}_{\rm pert}^{\gamma^* p}&=&\int^{P_\perp}_{\mu_b}\frac{d\mu}{\mu}\frac{\alpha_sC_A}{\pi}
\left[\ln\frac{P_\perp^2}{\mu^2}-2\beta_0+2\ln\frac{\hat{s}+Q^2}{\hat s}\right.\nonumber\\
&&\left.+~\frac{2C_F}{C_A} c_0^{\gamma g} 
\right]\ . \label{sudis}
\end{eqnarray}
and the nonperturbative part is the same as (\ref{npgp}). 

\begin{figure}[tbp]
\begin{center}
\includegraphics[width=7cm]{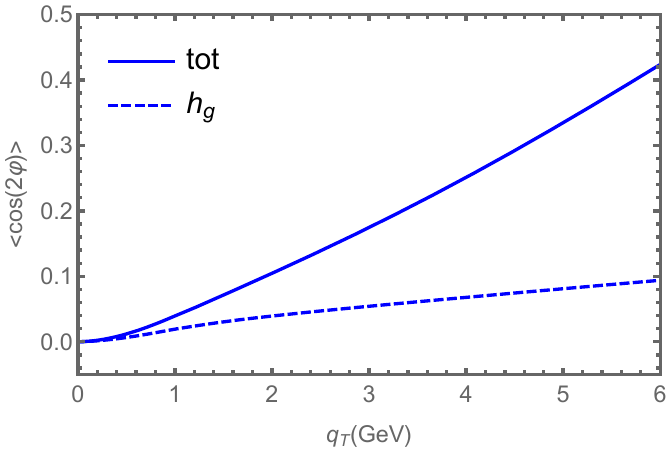}
\includegraphics[width=7cm]{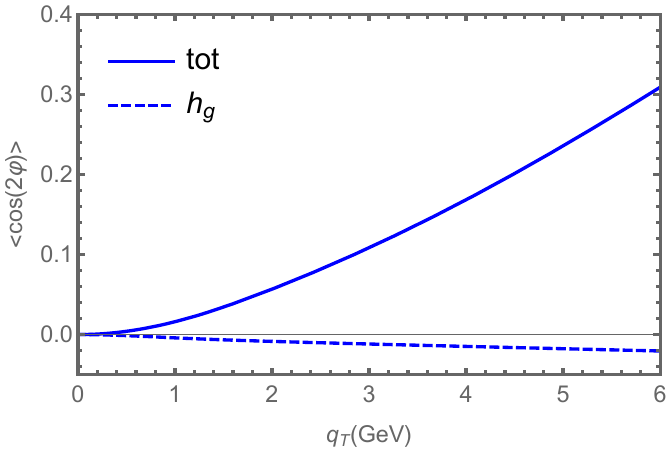}
\end{center}
\caption[*]{The $\cos 2 \phi$  azimuthal asymmetries for the di-jet  production   as the function of $q_\perp$  with $Q=10$  GeV, $P_\perp=15$ GeV, $z=\frac{1}{2}$, $\sqrt{s}=100 $  GeV  and $R=0.4$. 
The top and bottom panels are for the longitudinally and transversely polarized photons, respectively. The difference between the solid and dashed curves is due to the final state soft gluon emissions which contribute  equally in the longitudinal and transverse cases. }
\label{fig:cos2phiEIC}
\end{figure}

On the other hand, the $\cos(2\phi)$ term consists of two parts
\begin{eqnarray}
\widetilde{W}_2^{\gamma^* p}(b_\perp)=\frac{\alpha_sC_F}{\pi}c_2^{\gamma g}\widetilde{W}_0^{\gamma^* p}(b_\perp) +\frac{\sigma_2}{\sigma_0}\widetilde{W}_h^{\gamma^* p}(b_\perp)\ .
\end{eqnarray}
The first term comes from the soft gluon emission from the final state jets~\cite{Hatta:2020bgy}, and the second term comes from the linearly polarized gluon distribution whose resummation has been discussed in the previous subsection.\footnote{Our one-loop result only demonstrates the contribution from the linearly polarized gluon distribution. The associated Collins-Soper evolution for the linearly polarized gluon distribution will eventually lead to a resummation following the discussion in previous section. There is also a soft factor associated with the final state jet. We expect the same resummation formalism as that in $\widetilde{W}_0$.} Note that the latter is absent in photo-production studied in \cite{Hatta:2020bgy} and also in Section III  because $\sigma_2$ vanishes when $Q^2=0$, see (\ref{ratio}). 
All in all, we arrive at the following representation 
\begin{eqnarray}
 \widetilde{W}_2^{\gamma^* p}(b_\perp)&&=e^{-{\rm Sud}_{\rm pert}^{\gamma^* p}(b_*)-{\rm Sud}_{\rm NP}^{\gamma^*p}(b_\perp) }   \nonumber \\
&& \times  \Biggl[x_gf_g(x_g,\mu_b)\left(\frac{\alpha_sC_F}{\pi}c_2^{\gamma g}+\frac{\sigma_2}{\sigma_0}g_h(b_\perp)\right) \nonumber\\
&& + \frac{\sigma_2}{\sigma_0} \int\frac{dx'}{x'}{x_g}f_{i}(x',\mu)  C_{h/i}^{(1)}\left(\frac{x_g}{x'}\right) \Biggr]  \ . \label{three}
\end{eqnarray} 
Eq.~(\ref{three}) clearly exhibits the three distinct contributions mentioned at the beginning of this section: the intrinsic ($\sim g_h$) and radiative ($\sim C_{h/i}$) contributions to the linearly polarized gluon distribution, and the soft gluon emission contribution ($\sim c_2^{\gamma g}$). 

The numerical results for the $\cos 2\phi $ asymmetry  are  presented in Fig.~\ref{fig:cos2phiEIC} for  longitudinal (top panel) and transverse (bottom panel) virtual photons for a typical EIC kinematics with $R=0.4$. While the contribution from the linearly polarized gluon distribution (dashed curve) is noticeable, it is overwhelmed by that from the final state soft gluon emissions in the whole range of $q_\perp$.  The latter is independent of the polarization (longitudinal/transverse) of the virtual photon. In order to extract $h_g$ from this observable, it is probably better to use  larger values of $R$, say, $R=1$   to suppress the final state emissions. If one is ultimately interested in the `intrinsic' part of $h_g$,  further considerations are required (such as lowering $Q$) to suppress the `collinear radiative' part of $h_g$, see Fig.~\ref{fig:hgqkt}.

\section{Lepton Pair Production in Two-photon Process}

As a final example, we consider lepton pair production in QED $\gamma \gamma \to \ell^+(k_1) \ell^-(k_2)$ which has been actively studied recently in ultraperipheral collisions (UPC) at RHIC and the LHC \cite{Aaboud:2018eph,Adam:2018tdm,Lehner:2019amb,Adam:2019mby,ATLAS:2019vxg,Sirunyan:2020vvm,Aad:2020dur}. Similarly to the dijet problem in QCD, the dilepton azimuthal correlation  
is dominated by soft photon radiations from the final state leptons at small $q_\perp=k_1+k_2$,
\begin{equation}
|{\cal M}^{(1)}|^2_{\rm soft}=e^2\frac{2k_1\cdot k_2}{k_1\cdot k_s k_2\cdot k_s}|{\cal M}^{(0)}|^2 \ .
\end{equation}
where ${\cal M}^{(0)}$ represents the leading order Born amplitude and the soft photon carries momentum $k_s$.

Working in the laboratory frame, we integrate the soft emission kernel over the photon rapidity 
\begin{eqnarray}
  &&  \frac{\hat s}{P_\perp^2q_{\perp}^2}\int dy_g\frac{1}{\left(\sqrt{1+m_t^2}\cosh(y_g-y_2)+\cos(\phi)\right)}\nonumber\\
    &&~~~~\times \frac{1}{\left(\sqrt{1+m_t^2}\cosh(y_g-y_1)-\cos(\phi)\right)} \ ,\label{eq:sglepton0}
\end{eqnarray}
where $m_t^2\equiv m^2/P_\perp^2$ and $m$ is the lepton mass. Because of the mass term, there is no collinear divergence  associated with final state radiations. In other words, $m_t$ plays the role of $R$ in the previous sections. The integral is carried out in Appendix \ref{appmassive}, with the following result for the one-loop soft factor
\begin{equation}
S_{{\rm real}}(q_\perp)=\frac{\alpha}{\pi^2}\frac{1}{q_\perp^2}\ln\frac{Q^2}{m^2}+{\cal O}(m^2) \ ,\label{soft0}
\end{equation}
where $Q^2$ the invariant mass squared of the lepton pair.   Adding the virtual  contribution, we obtain the  soft factor in $b_\perp$-space   
\begin{eqnarray}
\widetilde{S}(b_\perp;Q,m)=-\frac{\alpha}{\pi}\ln\frac{Q^2}{m^2}\ln\frac{Q^2b_\perp^2}{c_0^2} \ .
\end{eqnarray}
Comments are in order regarding the relation to  Ref.~\cite{Klein:2020jom}. In this reference, the photon rapidity integral was carried out in the lepton frame (outgoing leptons are along the $z$-axis), not in the lab frame. This resulted in a different expression than (\ref{soft0}), and the two results lead to slightly different predictions for the acoplanarity of the lepton pair in UPC at the LHC, mainly in the moderate acoplanarity region. However, both predictions are compatible with the data due to  large  experimental uncertainties.

In addition to the angular independent part (\ref{soft0}), we have also calculated the  $\cos(2\phi)$ term in Appendix \ref{appmassive}.  In the small mass limit, the ratio $c_2/c_0$ is close to unity. This is because  soft photons are concentrated around the lepton directions due to the collinear enhancement. In the limit $m\to0$, the $\phi$-distribution of photons diverges  around $\phi=0$ and $\pi$.  
\begin{figure}[tbp]
\begin{center}
\includegraphics[width=7cm]{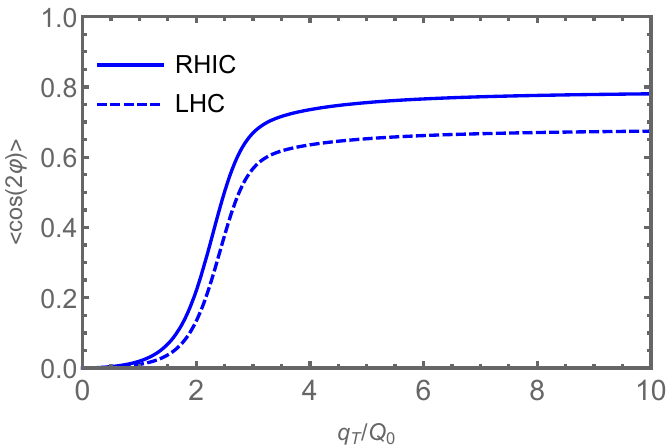}
\end{center}
\caption[*]{All order resummation result for the $\cos(2\phi)$ asymmetry in lepton pair production in two photon scattering process in the typical kinematics of UPC heavy ion collisions with the invariant mass of the muon pair $Q\approx 10~{\rm GeV}$ at LHC and the invariant mass of the electron pair $Q\approx 1~{\rm GeV}$ at RHIC, initial two photons contribute to a Gaussian distribution for the total transverse momentum with average transverse momentum $Q_0\approx 40~{\rm MeV}$. The asymmetry is shown as a  function of $q_\perp/Q_0$.}
\label{fig:leptonv2}
\end{figure}

All-order resummation can be carried out similarly to the previous problems. In place of the nonperturbative Sudakov form factor, we introduce a simple Gaussian factor which takes into account the intrinsic transverse momentum of incoming photons $\langle p_\perp\rangle$. The total transverse momentum distribution can be written as
\begin{eqnarray}
\frac{dN}{d^2q_\perp}&\propto &\int\frac{d^2b_\perp}{(2\pi)^2}e^{iq_\perp\cdot b_\perp} e^{-\frac{b_\perp^2 Q_0^2}{4}}e^{\widetilde{S}(Q,m;b_\perp)}\nonumber\\
&&~~\times\left[1-\beta c_{2\gamma} 2\cos(2\phi_b)+\cdots \right] \ ,\label{res0}
\end{eqnarray}
where $\beta=\frac{\alpha_e}{\pi} \ln\frac{Q^2}{m^2}$. For simplicity, let us set $y_1=y_2$ and define 
\begin{equation}
c_{2\gamma}=\frac{c_2}{c_0} = \frac{\ln \frac{1}{2m_t}}{\ln \frac{2}{m_t}}. \label{c2gamma}
\end{equation}
The general results provided in Appendix \ref{appmassive} can be used to compute cases with arbitrary $\Delta y$. We employ the value $Q_0=40\, {\rm MeV}\sim 1/R_A$ for the Gaussian width.
The Fourier transform leads to the following result
\begin{eqnarray}
&&\frac{dN}{d^2q_\perp}\propto\frac{\left(\frac{{Q_0^2}e^{-2\gamma_E}}{Q^2}\right)^\beta}{\pi Q_0^2} \left[\Gamma(1-\beta) {}_1F_1\left(1-\beta,1,-\frac{q_\perp^2}{Q_0^2}\right)\right.\nonumber\\
&&\left.+c_{2\gamma}2\cos(2\phi)\frac{q_\perp^2}{2Q_0^2}\Gamma(2-\beta){}_1F_1\left(2-\beta,3,-\frac{q_\perp^2}{Q_0^2}\right)\right]\ , \,\,\,
\end{eqnarray}
where ${}_1F_1$ is a hypergeometric function. We thus arrive at 
\begin{equation}
\langle \cos(2\phi)\rangle=c_{2\gamma}\beta\frac{(1-\beta)q_\perp^2}{2Q_0^2}\frac{{}_1F_1\left(2-\beta,3,-\frac{q_\perp^2}{Q_0^2}\right)}{{}_1F_1\left(1-\beta,1,-\frac{q_\perp^2}{Q_0^2}\right)} \ .
\end{equation}
For the typical kinematics of dimuon production from the two photon processes in UPC heavy ion collision at the LHC, $\beta\approx 0.02$ and $m/P_\perp\approx 0.02$ or $c_{2\gamma}\approx 0.70$. For RHIC kinematics, the typical values of $P_\perp$ range from 350 MeV to 2 GeV. Taking, for example, $Q=2P_\perp=1$ GeV ($y_1=y_2$), we get  $c_{2\gamma}\approx 0.82$ and $\beta \approx 0.035$ for di-electron production. 

In Fig.~\ref{fig:leptonv2}, we plot $\langle \cos(2\phi)\rangle$  as a function of  $q_\perp/Q_0$ for the above RHIC and LHC kinematics. We find that the asymmetry is very small when $q_\perp$ is below $2Q_0$. However, it rapidly grows around $2Q_0$ and becomes sizable at $3Q_0$ which is about $100~\rm MeV$. Similar results can be obtained for $\langle \cos(4\phi)\rangle$ as well. Both  the  $\cos 2\phi$ and $\cos 4\phi$ azimuthal asymmetries in di-lepton production  can also be induced by the  primordial  linearly polarized {\it photon} distribution~\cite{Li:2019yzy,Li:2019sin,Adam:2019mby}.  Our result shows that the high-$q_\perp$ tail of the asymmetries are overwhelmingly  developed via perturbative final state soft photon radiations. In contrast, at low $q_\perp$ the asymmetries are mainly attributed to the primordial linearly polarized photon distribution.

\section{Conclusions}

In summary, we have performed a systematic study of  azimuthal angular correlations between the total $\vec{q}_\perp=\vec{k}_{1\perp}+\vec{k}_{2\perp}$ and relative $\vec{P}_\perp=(\vec{k}_{1\perp}-\vec{k}_{2\perp})/2$ transverse momenta in dijet and related  systems. The resummation of the double and single logarithms $(\ln P_\perp/q_\perp)^n$ developed in \cite{Catani:2014qha,Catani:2017tuc,Hatta:2020bgy} is demonstrated for a number of processes. The dominant Fourier modes are $\cos (2\phi)$ and $\cos (\phi)$ for dijet and single jet (plus a color-neutral particle) productions, respectively. The expectation value $\langle \cos (n\phi)\rangle$ grows as $q_\perp^n$ in the small-$q_\perp$ region, and it can easily reach 10-20\% in the large-$q_\perp$ region. While this is an interesting feature of soft gluon emissions in its own right, it can become a serious background for certain purposes. In particular, the $\cos (2\phi)$ asymmetry in inclusive dijet production proposed as a signal of the linearly polarized gluon distribution \cite{Boer:2010zf,Metz:2011wb,Dumitru:2015gaa,Boer:2016fqd,Boer:2017xpy} inevitably faces this challenge. 

A majority of the processes we have studied are relevant to the future EIC experiment~\cite{Accardi:2012qut,Proceedings:2020eah,AbdulKhalek:2021gbh}, such as lepton-jet correlation in DIS (Sec. IIA), diffractive (Sec. IIIA) and inclusive (Sec. IIIB)  dijet production in photon-proton collisions, and inclusive dijet production in DIS (Sec. IV), see the plots in Figs.~\ref{fig:ep},\ref{fig:diffv2qtEIC},\ref{fig:gpdijetv2qt},\ref{fig:cos2phiEIC}. The comparative study of all these processes will provide a crucial test of our predictions and lead to a better constraint on the nucleon/nucleus tomography in terms of the gluon Wigner distribution and the linearly polarized gluon distribution. 

One of the important directions for future research is the understanding of power corrections of the form  $(q_\perp/P_\perp)^n$. We have heuristically  noticed in Section IIIA (see Fig.~\ref{fig:diffv2qt}) that the inclusion of power corrections in the soft emission kernel results in a better agreement with the CMS data in the large-$q_\perp$ region. The importance of power corrections in the hard part for the extraction of the linearly polarized gluon distribution has been discussed recently \cite{Altinoluk:2021ygv,Boussarie:2021lkb}. A combined analysis of resummation and power-corrections seems to be necessary to correctly interpret the experimental data.

\section*{Acknowledgments}  This material is based upon work supported by the U.S. Department of Energy, Office of Science, Office of Nuclear Physics, under contract numbers DE-AC02-05CH11231, DE- SC0012704, and within the framework of the TMD Topical Collaboration.  This work is also supported by the National Natural Science Foundations of China under Grant No. 11675093, No. 11575070, as well as by the CUHK-Shenzhen grant UDF01001859.

 \begin{widetext}

\appendix

\section{The evaluation of $S_g(k_J,p_1)$}
\label{s1}
In this appendix, we  evaluate the integral (\ref{int}) relevant to lepton-jet production. Introducing the rapidities $y_g$, $y_J$ of the soft gluon and the ougoing jet, respectively, we can write  
\begin{eqnarray}
\int\frac{dy_g}{2} S_g(k_J,p_1)&=& \int \frac{dy_g}{2} \frac{2k_J\cdot p_1}{k_J\cdot k_g \, p_1\cdot k_g} 
=  \frac{1}{k_{g\perp}^2}\int_{-\infty}^\Lambda d\Delta y \frac{e^{\Delta y}}{\cosh \Delta y -\cos \phi} \,,
\end{eqnarray}
where $\Delta y \equiv y_g -y_{J}$. $\Lambda$ is the rapidity cutoff which can be determined by the kinematical constraint $k_g^+ < p_1^+$, or more explicitly, 
\begin{equation}
k_{g\perp}e^{y_g} <p_1^+ \quad \rightarrow \quad e^{\Delta y} < \frac{p_1^+ e^{-y_J}}{k_{g\perp}}   \quad \rightarrow \quad \Delta y <\Lambda \equiv \frac{1}{2} \ln \frac{Q^4}{k_{g\perp}^2 k_{J\perp}^2},
\end{equation}
where $Q^2= -t =  -(k_1-p_1)^2 = p_1^+ k_{J\perp} e^{-y_J}$. Here we define $p^\pm \equiv p^0 \pm p^3$ and parametrize the momentum of a particle as $p^\mu = (p^+, p^-, p_\perp)$ where $p^\mu p_\mu= p^+p^- - p_\perp^2$. For massless particles, $p^\pm=p_\perp e^{\pm y}$.  After integrating over $\Delta y$ assuming $\Lambda \gg 1$, one finds
\begin{eqnarray}
\int_{-\infty}^\Lambda d\Delta y \frac{e^{\Delta y}}{\cosh \Delta y -\cos \phi} 
&=& \left . \frac{2\cos \phi}{\sin \phi} \tan^{-1} \left( \frac{e^{\Delta y} -\cos \phi }{\sin\phi }\right) \right|_{-\infty}^{\Delta y =\Lambda} + \left . \ln \left[1+e^{2\Delta y} -2e^{\Delta y} \cos \phi \right] \right|_{-\infty}^{\Delta y =\Lambda} \nonumber \\
&=& \frac{2\cos \phi}{\sin \phi}(\pi -\phi) + \ln \frac{Q^4}{k_{g\perp}^2 k_{J\perp}^2} + (\text{power corrections} \, \propto e^{-\Lambda}), \label{power}
\end{eqnarray}
where only the leading power contributions are kept. We can also implement the constraint (\ref{condition}) by restricting the $y_g$ integral for $R>|\phi|$ as 
\begin{equation}
\int_{-\infty}^{y_-}d\Delta y + \int_{y_+}^\Lambda d\Delta y\,,
\end{equation}
where $y_{\pm}=\pm\sqrt{R^2-\phi^2}$. The result can then be expanded in Fourier series
\begin{equation}
S_g(k_1,p_1) \Rightarrow \frac{1}{q_\perp^2}\left[\ln\frac{Q^2}{q_\perp^2}+\ln\frac{Q^2}{k_{J\perp}^2}+c_0+2c_1\cos(\phi)+2c_2\cos(2\phi) +\cdots\right]. \label{fou}
\end{equation}

\section{The evaluation of $S_g(k_1,k_2)$}
\label{s2}
Next we turn to the kernel relevant to dijet processes
\begin{eqnarray}
S_g(k_1,k_2)= \frac{2k_1\cdot k_2}{k_1\cdot k_g k_2\cdot k_g} \ .
\end{eqnarray}
where $k_{1,2}$ are the jet momenta. We can write 
\begin{equation}
  \int \frac{d\phi}{2\pi} \int dy_g S_g(k_1,k_2)=\frac{\hat s}{k_\perp^2k_{g\perp}^2}\int \frac{d\phi}{2\pi} \int dy_g\frac{\Theta (\Delta_{k_1k_g}>R^2) \Theta (\Delta_{k_2k_g}>R^2)}{\left(\cosh(y_g-y_2)+\cos \phi\right)\left(\cosh(y_g-y_1)-\cos \phi\right)}  \ ,
\end{equation}
where $\hat{s}=(k_1+k_2)^2$. We have inserted two kinematic constraints to exclude inside-jet radiations. Due to symmetries $\phi\to -\phi$, $\phi \to \pi-\phi$, we may restrict ourselves to the region $\pi/2> \phi\ge0$. The leading contribution depends on the rapidity difference $\Delta y_{12}=|y_1-y_2|$ between the two jets. To find the analytic expression for the above integral in the small-$R$ limit, we use the following  identities
\begin{eqnarray}
 \int_{-\infty}^{\infty}dy_g \frac{1+\cosh{\Delta y_{12}}}{2 (\cosh (y_g-y_2)+\cos{\phi})(\cosh (y_g-y_1)-\cos{\phi})} &=&\frac{\Delta y_{12} \sinh{\Delta y_{12}} + (\pi -2\phi)\cot{\phi} (\cosh{\Delta y_{12}}+1)}{\cosh{\Delta y_{12}}+\cos{2\phi}},\\
 \frac{2}{\pi} \int_0^{\pi/2} d\phi \frac{\Delta y_{12} \sinh{\Delta y_{12}}}{\cosh{\Delta y_{12}}+\cos{2\phi}}& =& \Delta y_{12}, \label{inte1} \\
  \frac{2}{\pi} \int_0^{\pi/2} d\phi (\pi -2\phi)\cot{\phi} \left[\frac{ \cosh{\Delta y_{12}}+1}{\cosh{\Delta y_{12}}+\cos{2\phi}} -1 \right] &=& -\Delta y_{12} + \ln{\left[2(1+\cosh{\Delta y_{12}})\right]}.\label{inte2}
\end{eqnarray}
As for the angular independent contribution, we  find 
\begin{equation}
    c_0= \ln \frac{1}{R^2} + \ln{\left[2(1+\cosh{\Delta y_{12}})\right]} = \ln \frac{a_0}{R^2}, \label{c0cone}
\end{equation}
where $\ln \frac{1}{R^2}$ arises from the incomplete cancellation between the full space integration and the region inside the jet cone when the singular term $\frac{2}{\pi} \int d\phi (\pi -2\phi)\cot{\phi}$ is evaluated. Similarly, by using the same identities (\ref{inte1}) and  (\ref{inte2}), we can obtain
\begin{equation}
    c_2= \ln \frac{1}{R^2} + \Delta y_{12} \sinh{\Delta y_{12}} - \cosh{\Delta y_{12}} \ln\left[2\left(1+\cosh{\Delta y_{12}}\right)\right]= \ln \frac{a_2}{R^2}. \label{c2cone}
\end{equation} 
The result simplifies in two limits  
\begin{equation}
c_2=\ln \frac{1}{R^2}-\ln 4,
\end{equation}
for $y_1=y_2$, and 
\begin{equation}
c_2 = \ln \frac{1}{R^2}-1,
\end{equation}
for $|y_1-y_2|\gg 1$. 

The above results can be readily generalized to all possible `dipole' radiators (also known as the eikonal factors) in $2\to 2$ processes $a(p_1)+b(p_2)\to c(k_1)+d(k_2)$ with four on-shell external massless particles as follows 
\begin{eqnarray}
S_g(p_1,p_2)  &\Rightarrow& \frac{1}{q_\perp^2}\left[2 \ln\frac{\hat s}{q_\perp^2}\right],  \\ 
S_g(k_1,p_1) &\Rightarrow& \frac{1}{q_\perp^2}\left[\ln\frac{\hat s}{q_\perp^2}+\ln\frac{\hat t}{\hat u}+c_0^{\text{fi}}+2\sum_{n=1}^{\infty}c_n^{\text{fi}}\cos(n\phi) (-1)^n\right],  \\
S_g(k_2,p_1) &\Rightarrow& \frac{1}{q_\perp^2}\left[\ln\frac{\hat s}{q_\perp^2}+\ln\frac{\hat u}{\hat t}+c_0^{\text{fi}}+2\sum_{n=1}^{\infty}c_n^{\text{fi}}\cos(n\phi)\right], \\
S_g(k_1,p_2) &\Rightarrow& \frac{1}{q_\perp^2}\left[\ln\frac{\hat s}{q_\perp^2}+\ln\frac{\hat u}{\hat t}+c_0^{\text{fi}}+2\sum_{n=1}^{\infty}c_n^{\textrm{fi}}\cos(n\phi) (-1)^n\right], \\
S_g(k_2,p_2) &\Rightarrow& \frac{1}{q_\perp^2}\left[\ln\frac{\hat s}{q_\perp^2}+\ln\frac{\hat t}{\hat u}+c_0^{\text{fi}}+2\sum_{n=1}^{\infty}c_n^{\text{fi}}\cos(n\phi)\right],\label{dis} \\
S_g(k_1,k_2) &\Rightarrow& \frac{2}{q_\perp^2}\left[\ln\frac{\hat s}{-\hat u}+\ln\frac{\hat s}{-\hat t}+c_0^{\text{ff}}+2\sum_{n=1}^{\infty}c_{2n}^{\text{ff}}\cos(2n\phi)\right], 
\end{eqnarray}
where we find in the small cone limit $c_0^{\text{fi}}= \ln 1/R^2$, $c_n^{\text{fi}}= \ln 1/R^2 +f(n)$, $c_0^{\text{ff}} =\ln 1/R^2$, and 
\begin{equation}
c_2^{\text{ff}}=  \ln \frac{1}{R^2} + \frac{\hat u}{\hat t} \ln \frac{-\hat u}{\hat s}+\frac{\hat t}{\hat u} \ln \frac{-\hat t}{\hat s}= \ln \frac{a_2}{R^2}.
\end{equation}
It is clear that $S_g(p_1,p_2)$ is independent of azimuthal angle since the pure initial state gluon radiation is expected to be symmetric. In general, there are two types of anisotropy generated from the final state radiations, as shown above. The eikonal factors $S_g (k_i, p_j)$ involve one final-state jet and one initial-state particle, and their contributions are captured by the coefficients $c_n^{\text{fi}}$. For odd Fourier coefficients, there is a sign change between $S_g(k_1, p_i)$ and $S_g(k_2, p_i)$. This is due to the fact that the final state gluon radiation is favored along the jet direction and it contributes to odd coefficients oppositely (as shown in Fig.~\ref{fig:dijet0} $\cos (\pi -\phi) =-\cos \phi$) for two jets that are back-to-back. Besides, more complicated angular correlations characterized by $c_n^{\text{ff}}$ can arise from $S_g(k_1,k_2)$ that depends on both final-state jets. Note that in $e+q\to e'+jet$ scattering in DIS, $\hat{s}\hat{t}/\hat{u}=Q^4/k_{J\perp}^2$ so (\ref{dis}) reduces to (\ref{fou}) when $p_2$ and $k_2$ in Eq.~(\ref{dis}) are identified as the incoming and outgoing quark momenta, respectively. Together with the usual factor of $\alpha_s/(2\pi^2)$ and the corresponding color factor, the above identities give rise to the one-loop results for various scattering processes.

\section{ $S_g(k_1,k_2)$ for massive  final state particles}
\label{appmassive}

In this appendix we again consider $S_g(k_1,k_2)$, but now we assume that the final state particles are  massive with mass $m$. We are interested in the regime  $m_t\equiv m/k_\perp\ll 1$ where $k_\perp=|k_{1\perp}|=|k_{2\perp}|$. The relevant integral is  
\begin{eqnarray}
I(\phi)&\equiv&  \int_{-\infty}^\infty dy_g\frac{1+\cosh \Delta y_{12}}{2(A\cosh (y_1 -y_g)-\cos \phi)(A\cosh (y_2-y_g)+\cos \phi)} \nn
&=&\frac{1+\cosh \Delta y_{12}}{4A\cosh \delta} \int \frac{dy_g}{\cosh y_g}  \left(\frac{1}{A\cosh (y_g-\delta)-\cos \phi} +\frac{1}{A\cosh (y_g+\delta)+\cos \phi} \right)
\end{eqnarray}
where $A\equiv \sqrt{1+m_t^2}\ge 1$ and $\delta = |y_1-y_2|/2=\Delta y_{12}/2$.  The collinear singularity at $\phi=0$ is regularized by $m_t\neq 0$, so there is no need to impose kinematical constraints. 
We change variables as $e^{y_g}=z$ and $e^{-\delta}=c$ and obtain
\begin{eqnarray}
I&=& \frac{1+\cosh \Delta y_{12}}{A\cosh\delta}\int_0^\infty \frac{dz}{1+z^2}\left(\frac{cz}{A(1+c^2z^2)-2cz \cos \phi} + (c\to 1/c,\ \phi\to \pi-\phi)\right) \nn 
&=&\frac{2(1+\cosh\Delta y_{12})}{2A^2\sinh^2 \delta +\cos 2\phi +1}\Biggl[\tan^{-1} \left(\frac{\cos\phi}{\sqrt{A^2-\cos^2\phi}}\right) \frac{ \cos \phi}{\sqrt{A^2-\cos^2\phi}} + \delta\tanh\delta \Biggr] \nn 
&\approx& \frac{1+\cosh \Delta y_{12}}{\cosh \Delta y_{12} +\cos 2\phi }\Biggl[\left(\pi-2\phi\right) \frac{ \cos \phi}{\sqrt{\sin^2\phi + m_t^2}} +  \frac{\Delta y_{12}\sinh \Delta y_{12}}{\cosh \Delta y_{12}+1}\Biggr] ,
\end{eqnarray}
where in the last expression we have set $m_t=0$ wherever it is safe to do so. 
Integrating over $\phi$ with the help of (\ref{inte1}), (\ref{inte2}), we find 
\beq
\frac{1}{2\pi}\int_0^{2\pi} d\phi I(\phi)=\frac{4}{2\pi}\int_0^{\pi/2} d\phi I(\phi) &=&\ln \frac{2(1+\cosh \Delta y_{12})}{m_t^2}  +{\cal O}(m_t^2) \nn 
&=& \ln \frac{Q^2}{m^2}+{\cal O}(m^2),
\eeq
where $Q^2$ is the invariant mass of the lepton pair. This result was previously obtained in \cite{Sun:2015doa} in a different way. Similarly, for the coefficient of $\cos 2\phi$, we get 
\beq
\frac{4}{2\pi}\int_0^{\pi/2} d\phi \cos 2\phi I(\phi) &=& \ln \frac{1}{m_t^2}+\Delta y_{12}\sinh \Delta y_{12} -\cosh \Delta y_{12} \ln[2(1+\cosh \Delta y_{12})] + {\cal O}(m^2) \nn 
&=& \ln \frac{Q^2}{m^2} +g(\Delta y_{12})
+ {\cal O}(m^2).
\eeq
where 
\beq
g(y)= y\sinh y -2\cosh^2 \frac{y}{2} \ln[2(1+\cosh y)]
\eeq
is always negative and satisfies $g(0)=-2\ln 4\approx -2.77$ and $g(y)\approx -y$ as $y\to \infty$. Note that these results are  identical to (\ref{c0cone}) and (\ref{c2cone})  after  replacing $R\to m_t$.

\end{widetext}


\begin{thebibliography}{99}


\bibitem{Abazov:2004hm}
V.~M.~Abazov \textit{et al.} [D0],
Phys. Rev. Lett. \textbf{94}, 221801 (2005)
doi:10.1103/PhysRevLett.94.221801
[arXiv:hep-ex/0409040 [hep-ex]].

\bibitem{Abelev:2007ii}
B.~I.~Abelev \textit{et al.} [STAR],
Phys. Rev. Lett. \textbf{99}, 142003 (2007)
doi:10.1103/PhysRevLett.99.142003
[arXiv:0705.4629 [hep-ex]].

\bibitem{Khachatryan:2011zj}
V.~Khachatryan \textit{et al.} [CMS],
Phys. Rev. Lett. \textbf{106}, 122003 (2011)
doi:10.1103/PhysRevLett.106.122003
[arXiv:1101.5029 [hep-ex]].

\bibitem{daCosta:2011ni}
G.~Aad \textit{et al.} [ATLAS],
Phys. Rev. Lett. \textbf{106}, 172002 (2011)
doi:10.1103/PhysRevLett.106.172002
[arXiv:1102.2696 [hep-ex]].

\bibitem{Aad:2010bu}
G.~Aad \textit{et al.} [ATLAS],
Phys. Rev. Lett. \textbf{105}, 252303 (2010)
doi:10.1103/PhysRevLett.105.252303
[arXiv:1011.6182 [hep-ex]].

\bibitem{Chatrchyan:2011sx}
S.~Chatrchyan \textit{et al.} [CMS],
Phys. Rev. C \textbf{84}, 024906 (2011)
doi:10.1103/PhysRevC.84.024906
[arXiv:1102.1957 [nucl-ex]].

\bibitem{Adamczyk:2013jei}
L.~Adamczyk \textit{et al.} [STAR],
Phys. Rev. Lett. \textbf{112}, no.12, 122301 (2014)
doi:10.1103/PhysRevLett.112.122301
[arXiv:1302.6184 [nucl-ex]].

\bibitem{Aaboud:2019oop}
M.~Aaboud \textit{et al.} [ATLAS],
Phys. Rev. C \textbf{100}, no.3, 034903 (2019)
doi:10.1103/PhysRevC.100.034903
[arXiv:1901.10440 [nucl-ex]].

\bibitem{Accardi:2012qut}
A.~Accardi, J.~L.~Albacete, M.~Anselmino, N.~Armesto, E.~C.~Aschenauer, A.~Bacchetta, D.~Boer, W.~K.~Brooks, T.~Burton and N.~B.~Chang, \textit{et al.}
Eur. Phys. J. A \textbf{52}, no.9, 268 (2016)
doi:10.1140/epja/i2016-16268-9
[arXiv:1212.1701 [nucl-ex]].

\bibitem{Proceedings:2020eah}
Y.~Hatta, Y.~V.~Kovchegov, C.~Marquet, A.~Prokudin, E.~Aschenauer, H.~Avakian, A.~Bacchetta, D.~Boer, G.~A.~Chirilli and A.~Dumitru, \textit{et al.}
doi:10.1142/11684
[arXiv:2002.12333 [hep-ph]].

\bibitem{AbdulKhalek:2021gbh}
R.~Abdul Khalek, A.~Accardi, J.~Adam, D.~Adamiak, W.~Akers, M.~Albaladejo, A.~Al-bataineh, M.~G.~Alexeev, F.~Ameli and P.~Antonioli, \textit{et al.}
[arXiv:2103.05419 [physics.ins-det]].

\bibitem{Hatta:2016dxp}
Y.~Hatta, B.~W.~Xiao and F.~Yuan,
Phys. Rev. Lett. \textbf{116}, no.20, 202301 (2016)
doi:10.1103/PhysRevLett.116.202301
[arXiv:1601.01585 [hep-ph]].

\bibitem{Altinoluk:2015dpi}
T.~Altinoluk, N.~Armesto, G.~Beuf and A.~H.~Rezaeian,
Phys. Lett. B \textbf{758}, 373-383 (2016)
doi:10.1016/j.physletb.2016.05.032
[arXiv:1511.07452 [hep-ph]].

\bibitem{Zhou:2016rnt}
J.~Zhou,
Phys. Rev. D \textbf{94}, no.11, 114017 (2016)
doi:10.1103/PhysRevD.94.114017
[arXiv:1611.02397 [hep-ph]].

\bibitem{Hagiwara:2017fye}
Y.~Hagiwara, Y.~Hatta, R.~Pasechnik, M.~Tasevsky and O.~Teryaev,
Phys. Rev. D \textbf{96}, no.3, 034009 (2017)
doi:10.1103/PhysRevD.96.034009
[arXiv:1706.01765 [hep-ph]].

\bibitem{Mantysaari:2019csc}
H.~M\"antysaari, N.~Mueller and B.~Schenke,
Phys. Rev. D \textbf{99}, no.7, 074004 (2019)
doi:10.1103/PhysRevD.99.074004
[arXiv:1902.05087 [hep-ph]].

\bibitem{Mantysaari:2019hkq}
H.~M\"antysaari, N.~Mueller, F.~Salazar and B.~Schenke,
Phys. Rev. Lett. \textbf{124}, no.11, 112301 (2020)
doi:10.1103/PhysRevLett.124.112301
[arXiv:1912.05586 [nucl-th]].

\bibitem{Boer:2010zf}
D.~Boer, S.~J.~Brodsky, P.~J.~Mulders and C.~Pisano,
Phys. Rev. Lett. \textbf{106}, 132001 (2011)
doi:10.1103/PhysRevLett.106.132001
[arXiv:1011.4225 [hep-ph]].

\bibitem{Metz:2011wb}
A.~Metz and J.~Zhou,
Phys. Rev. D \textbf{84}, 051503 (2011)
doi:10.1103/PhysRevD.84.051503
[arXiv:1105.1991 [hep-ph]].

\bibitem{Dumitru:2015gaa}
A.~Dumitru, T.~Lappi and V.~Skokov,
Phys. Rev. Lett. \textbf{115}, no.25, 252301 (2015)
doi:10.1103/PhysRevLett.115.252301
[arXiv:1508.04438 [hep-ph]].

\bibitem{Boer:2017xpy}
D.~Boer, P.~J.~Mulders, J.~Zhou and Y.~j.~Zhou,
JHEP \textbf{10}, 196 (2017)
doi:10.1007/JHEP10(2017)196
[arXiv:1702.08195 [hep-ph]].

\bibitem{Boer:2016fqd}
D.~Boer, P.~J.~Mulders, C.~Pisano and J.~Zhou,
JHEP \textbf{08}, 001 (2016)
doi:10.1007/JHEP08(2016)001
[arXiv:1605.07934 [hep-ph]].

\bibitem{Xing:2020hwh}
H.~Xing, C.~Zhang, J.~Zhou and Y.~J.~Zhou,
JHEP \textbf{10}, 064 (2020)
doi:10.1007/JHEP10(2020)064
[arXiv:2006.06206 [hep-ph]].

\bibitem{Zhao:2021kae}
Y.~Y.~Zhao, M.~M.~Xu, L.~Z.~Chen, D.~H.~Zhang and Y.~F.~Wu,
[arXiv:2105.08818 [hep-ph]].

\bibitem{Banfi:2003jj}
A.~Banfi and M.~Dasgupta,
JHEP \textbf{01}, 027 (2004)
doi:10.1088/1126-6708/2004/01/027
[arXiv:hep-ph/0312108 [hep-ph]].

\bibitem{Banfi:2008qs}
A.~Banfi, M.~Dasgupta and Y.~Delenda,
Phys. Lett. B \textbf{665}, 86-91 (2008)
doi:10.1016/j.physletb.2008.05.065
[arXiv:0804.3786 [hep-ph]].

\bibitem{Hautmann:2008vd}
F.~Hautmann and H.~Jung,
JHEP \textbf{10}, 113 (2008)
doi:10.1088/1126-6708/2008/10/113
[arXiv:0805.1049 [hep-ph]].

\bibitem{Mueller:2013wwa}
A.~H.~Mueller, B.~W.~Xiao and F.~Yuan,
Phys. Rev. D \textbf{88}, no.11, 114010 (2013)
doi:10.1103/PhysRevD.88.114010
[arXiv:1308.2993 [hep-ph]].

\bibitem{Sun:2014gfa}
P.~Sun, C.~P.~Yuan and F.~Yuan,
Phys. Rev. Lett. \textbf{113}, no.23, 232001 (2014)
doi:10.1103/PhysRevLett.113.232001
[arXiv:1405.1105 [hep-ph]].

\bibitem{Sun:2015doa}
P.~Sun, C.~P.~Yuan and F.~Yuan,
Phys. Rev. D \textbf{92}, no.9, 094007 (2015)
doi:10.1103/PhysRevD.92.094007
[arXiv:1506.06170 [hep-ph]].

\bibitem{Hatta:2019ixj}
Y.~Hatta, N.~Mueller, T.~Ueda and F.~Yuan,
Phys. Lett. B \textbf{802}, 135211 (2020)
doi:10.1016/j.physletb.2020.135211
[arXiv:1907.09491 [hep-ph]].

\bibitem{Liu:2018trl}
X.~Liu, F.~Ringer, W.~Vogelsang and F.~Yuan,
Phys. Rev. Lett. \textbf{122}, no.19, 192003 (2019)
doi:10.1103/PhysRevLett.122.192003
[arXiv:1812.08077 [hep-ph]].

\bibitem{Liu:2020dct}
X.~Liu, F.~Ringer, W.~Vogelsang and F.~Yuan,
Phys. Rev. D \textbf{102}, no.9, 094022 (2020)
doi:10.1103/PhysRevD.102.094022
[arXiv:2007.12866 [hep-ph]].

\bibitem{Chien:2019gyf}
Y.~T.~Chien, D.~Y.~Shao and B.~Wu,
JHEP \textbf{11}, 025 (2019)
doi:10.1007/JHEP11(2019)025
[arXiv:1905.01335 [hep-ph]].

\bibitem{Chien:2020hzh}
Y.~T.~Chien, R.~Rahn, S.~Schrijnder van Velzen, D.~Y.~Shao, W.~J.~Waalewijn and B.~Wu,
Phys. Lett. B \textbf{815}, 136124 (2021)
doi:10.1016/j.physletb.2021.136124
[arXiv:2005.12279 [hep-ph]].

\bibitem{Catani:2014qha}
S.~Catani, M.~Grazzini and A.~Torre,
Nucl. Phys. B \textbf{890}, 518-538 (2014)
doi:10.1016/j.nuclphysb.2014.11.019
[arXiv:1408.4564 [hep-ph]].

\bibitem{Catani:2017tuc}
S.~Catani, M.~Grazzini and H.~Sargsyan,
JHEP \textbf{06}, 017 (2017)
doi:10.1007/JHEP06(2017)017
[arXiv:1703.08468 [hep-ph]].

\bibitem{Hatta:2020bgy}
Y.~Hatta, B.~W.~Xiao, F.~Yuan and J.~Zhou,
Phys. Rev. Lett. \textbf{126}, no.14, 142001 (2021)
doi:10.1103/PhysRevLett.126.142001
[arXiv:2010.10774 [hep-ph]].

\bibitem{CMS:2020ekd}
 [CMS],
CMS-PAS-HIN-18-011.

\bibitem{Chen:2020adz}
H.~Chen, I.~Moult and H.~X.~Zhu,
Phys. Rev. Lett. \textbf{126}, no.11, 112003 (2021)
doi:10.1103/PhysRevLett.126.112003
[arXiv:2011.02492 [hep-ph]].

\bibitem{Karlberg:2021kwr}
A.~Karlberg, G.~P.~Salam, L.~Scyboz and R.~Verheyen,
[arXiv:2103.16526 [hep-ph]].

\bibitem{Chen:2021gdk}
H.~Chen, I.~Moult and H.~X.~Zhu,
[arXiv:2104.00009 [hep-ph]].

\bibitem{Dasgupta:2002bw}
M.~Dasgupta and G.~P.~Salam,
JHEP \textbf{03}, 017 (2002)
doi:10.1088/1126-6708/2002/03/017
[arXiv:hep-ph/0203009 [hep-ph]].

\bibitem{Boer:2006eq}
D.~Boer and W.~Vogelsang,
Phys. Rev. D \textbf{74}, 014004 (2006)
doi:10.1103/PhysRevD.74.014004
[arXiv:hep-ph/0604177 [hep-ph]].

\bibitem{Berger:2007si}
E.~L.~Berger, J.~W.~Qiu and R.~A.~Rodriguez-Pedraza,
Phys. Lett. B \textbf{656}, 74-78 (2007)
doi:10.1016/j.physletb.2007.09.008
[arXiv:0707.3150 [hep-ph]].

\bibitem{Bacchetta:2008xw}
A.~Bacchetta, D.~Boer, M.~Diehl and P.~J.~Mulders,
JHEP \textbf{08}, 023 (2008)
doi:10.1088/1126-6708/2008/08/023
[arXiv:0803.0227 [hep-ph]].

\bibitem{Bacchetta:2019qkv}
A.~Bacchetta, G.~Bozzi, M.~G.~Echevarria, C.~Pisano, A.~Prokudin and M.~Radici,
Phys. Lett. B \textbf{797}, 134850 (2019)
doi:10.1016/j.physletb.2019.134850
[arXiv:1906.07037 [hep-ph]].

\bibitem{Nadolsky:2007ba}
P.~M.~Nadolsky, C.~Balazs, E.~L.~Berger and C.~P.~Yuan,
Phys. Rev. D \textbf{76}, 013008 (2007)
doi:10.1103/PhysRevD.76.013008
[arXiv:hep-ph/0702003 [hep-ph]].

\bibitem{Catani:2010pd}
S.~Catani and M.~Grazzini,
Nucl. Phys. B \textbf{845}, 297-323 (2011)
doi:10.1016/j.nuclphysb.2010.12.007
[arXiv:1011.3918 [hep-ph]].

\bibitem{Balitsky:2017flc}
I.~Balitsky and A.~Tarasov,
JHEP \textbf{07}, 095 (2017)
doi:10.1007/JHEP07(2017)095
[arXiv:1706.01415 [hep-ph]].

\bibitem{Balitsky:2017gis}
I.~Balitsky and A.~Tarasov,
JHEP \textbf{05}, 150 (2018)
doi:10.1007/JHEP05(2018)150
[arXiv:1712.09389 [hep-ph]].

\bibitem{Ebert:2018gsn}
M.~A.~Ebert, I.~Moult, I.~W.~Stewart, F.~J.~Tackmann, G.~Vita and H.~X.~Zhu,
JHEP \textbf{04}, 123 (2019)
doi:10.1007/JHEP04(2019)123
[arXiv:1812.08189 [hep-ph]].

\bibitem{Moult:2019mog}
I.~Moult, I.~W.~Stewart and G.~Vita,
JHEP \textbf{11}, 153 (2019)
doi:10.1007/JHEP11(2019)153
[arXiv:1905.07411 [hep-ph]].

\bibitem{Bertulani:1987tz}
C.~A.~Bertulani and G.~Baur,
Phys. Rept. \textbf{163}, 299 (1988)
doi:10.1016/0370-1573(88)90142-1

\bibitem{Adams:2004rz}
J.~Adams \textit{et al.} [STAR],
Phys. Rev. C \textbf{70}, 031902 (2004)
doi:10.1103/PhysRevC.70.031902
[arXiv:nucl-ex/0404012 [nucl-ex]].

\bibitem{Baur:2003ar}
G.~Baur, K.~Hencken, A.~Aste, D.~Trautmann and S.~R.~Klein,
Nucl. Phys. A \textbf{729}, 787-808 (2003)
doi:10.1016/j.nuclphysa.2003.09.006
[arXiv:nucl-th/0307031 [nucl-th]].

\bibitem{Hencken:2004td}
K.~Hencken, G.~Baur and D.~Trautmann,
Phys. Rev. C \textbf{69}, 054902 (2004)
doi:10.1103/PhysRevC.69.054902
[arXiv:nucl-th/0402061 [nucl-th]].

\bibitem{Baur:2007fv}
G.~Baur, K.~Hencken and D.~Trautmann,
Phys. Rept. \textbf{453}, 1-27 (2007)
doi:10.1016/j.physrep.2007.09.002
[arXiv:0706.0654 [nucl-th]].

\bibitem{Bertulani:2005ru}
C.~A.~Bertulani, S.~R.~Klein and J.~Nystrand,
Ann. Rev. Nucl. Part. Sci. \textbf{55}, 271-310 (2005)
doi:10.1146/annurev.nucl.55.090704.151526
[arXiv:nucl-ex/0502005 [nucl-ex]].

\bibitem{Baltz:2007kq}
A.~J.~Baltz, G.~Baur, D.~d'Enterria, L.~Frankfurt, F.~Gelis, V.~Guzey, K.~Hencken, Y.~Kharlov, M.~Klasen and S.~R.~Klein, \textit{et al.}
Phys. Rept. \textbf{458}, 1-171 (2008)
doi:10.1016/j.physrep.2007.12.001
[arXiv:0706.3356 [nucl-ex]].

\bibitem{Baltz:2009jk}
A.~J.~Baltz, Y.~Gorbunov, S.~R.~Klein and J.~Nystrand,
Phys. Rev. C \textbf{80}, 044902 (2009)
doi:10.1103/PhysRevC.80.044902
[arXiv:0907.1214 [nucl-ex]].

\bibitem{ATLAS:2016vdy}
 [ATLAS],
ATLAS-CONF-2016-025.

\bibitem{Klein:2018cjh}
S.~R.~Klein,
Phys. Rev. C \textbf{97}, no.5, 054903 (2018)
doi:10.1103/PhysRevC.97.054903
[arXiv:1801.04320 [nucl-th]].

\bibitem{CMS:2020avp}
 [CMS],
CMS-PAS-HIN-19-014.

\bibitem{Klein:2020fmr}
S.~Klein and P.~Steinberg,
Ann. Rev. Nucl. Part. Sci. \textbf{70}, 323-354 (2020)
doi:10.1146/annurev-nucl-030320-033923
[arXiv:2005.01872 [nucl-ex]].

\bibitem{Aaboud:2018eph}
M.~Aaboud \textit{et al.} [ATLAS],
Phys. Rev. Lett. \textbf{121}, no.21, 212301 (2018)
doi:10.1103/PhysRevLett.121.212301
[arXiv:1806.08708 [nucl-ex]].

\bibitem{Adam:2018tdm}
J.~Adam \textit{et al.} [STAR],
Phys. Rev. Lett. \textbf{121}, no.13, 132301 (2018)
doi:10.1103/PhysRevLett.121.132301
[arXiv:1806.02295 [hep-ex]].

\bibitem{Lehner:2019amb}
S.~Lehner [ALICE],
PoS \textbf{LHCP2019}, 164 (2019)
doi:10.22323/1.350.0164
[arXiv:1909.02508 [nucl-ex]].

\bibitem{Adam:2019mby}
J.~Adam \textit{et al.} [STAR],
[arXiv:1910.12400 [nucl-ex]].

\bibitem{ATLAS:2019vxg}
 [ATLAS],
ATLAS-CONF-2019-051.

\bibitem{Sirunyan:2020vvm}
A.~M.~Sirunyan \textit{et al.} [CMS],
[arXiv:2011.05239 [hep-ex]].

\bibitem{Aad:2020dur}
G.~Aad \textit{et al.} [ATLAS],
[arXiv:2011.12211 [nucl-ex]].

\bibitem{Klusek-Gawenda:2018zfz}
M.~K\l{}usek-Gawenda, R.~Rapp, W.~Sch\"afer and A.~Szczurek,
Phys. Lett. B \textbf{790}, 339-344 (2019)
doi:10.1016/j.physletb.2019.01.035
[arXiv:1809.07049 [nucl-th]].

\bibitem{Klein:2018fmp}
S.~Klein, A.~H.~Mueller, B.~W.~Xiao and F.~Yuan,
Phys. Rev. Lett. \textbf{122}, no.13, 132301 (2019)
doi:10.1103/PhysRevLett.122.132301
[arXiv:1811.05519 [hep-ph]].

\bibitem{Zha:2018tlq}
W.~Zha, J.~D.~Brandenburg, Z.~Tang and Z.~Xu,
Phys. Lett. B \textbf{800}, 135089 (2020)
doi:10.1016/j.physletb.2019.135089
[arXiv:1812.02820 [nucl-th]].

\bibitem{Li:2019yzy}
C.~Li, J.~Zhou and Y.~J.~Zhou,
Phys. Lett. B \textbf{795}, 576-580 (2019)
doi:10.1016/j.physletb.2019.07.005
[arXiv:1903.10084 [hep-ph]].

\bibitem{Li:2019sin}
C.~Li, J.~Zhou and Y.~J.~Zhou,
Phys. Rev. D \textbf{101}, no.3, 034015 (2020)
doi:10.1103/PhysRevD.101.034015
[arXiv:1911.00237 [hep-ph]].

\bibitem{Zhao:2019hta}
J.~Zhao and F.~Wang,
Prog. Part. Nucl. Phys. \textbf{107}, 200-236 (2019)
doi:10.1016/j.ppnp.2019.05.001
[arXiv:1906.11413 [nucl-ex]].

\bibitem{Karadag:2019gvc}
S.~Karada\u{g} and M.~C.~G\"u\c{c}l\"u,
Phys. Rev. C \textbf{102}, no.1, 014904 (2020)
doi:10.1103/PhysRevC.102.014904
[arXiv:1911.08507 [hep-ph]].

\bibitem{Vidovic:1992ik}
M.~Vidovic, M.~Greiner, C.~Best and G.~Soff,
Phys. Rev. C \textbf{47}, 2308-2319 (1993)
doi:10.1103/PhysRevC.47.2308

\bibitem{Klein:2020jom}
S.~Klein, A.~H.~Mueller, B.~W.~Xiao and F.~Yuan,
Phys. Rev. D \textbf{102}, no.9, 094013 (2020)
doi:10.1103/PhysRevD.102.094013
[arXiv:2003.02947 [hep-ph]].

\bibitem{Xiao:2020ddm}
B.~W.~Xiao, F.~Yuan and J.~Zhou,
Phys. Rev. Lett. \textbf{125}, no.23, 232301 (2020)
doi:10.1103/PhysRevLett.125.232301
[arXiv:2003.06352 [hep-ph]].

\bibitem{Klusek-Gawenda:2020eja}
M.~K\l{}usek-Gawenda, W.~Sch\"afer and A.~Szczurek,
Phys. Lett. B \textbf{814}, 136114 (2021)
doi:10.1016/j.physletb.2021.136114
[arXiv:2012.11973 [hep-ph]].

\bibitem{Brandenburg:2021lnj}
J.~D.~Brandenburg, W.~Zha and Z.~Xu,
[arXiv:2103.16623 [hep-ph]].



  
\bibitem{Gutierrez-Reyes:2018qez}
D.~Gutierrez-Reyes, I.~Scimemi, W.~J.~Waalewijn and L.~Zoppi,
Phys. Rev. Lett. \textbf{121}, no.16, 162001 (2018)
doi:10.1103/PhysRevLett.121.162001
[arXiv:1807.07573 [hep-ph]].

\bibitem{Gutierrez-Reyes:2019vbx}
D.~Gutierrez-Reyes, I.~Scimemi, W.~J.~Waalewijn and L.~Zoppi,
JHEP \textbf{10}, 031 (2019)
doi:10.1007/JHEP10(2019)031
[arXiv:1904.04259 [hep-ph]].

\bibitem{Arratia:2019vju}
M.~Arratia, Y.~Song, F.~Ringer and B.~V.~Jacak,
Phys. Rev. C \textbf{101}, no.6, 065204 (2020)
doi:10.1103/PhysRevC.101.065204
[arXiv:1912.05931 [nucl-ex]].

\bibitem{Arratia:2020nxw}
M.~Arratia, Z.~B.~Kang, A.~Prokudin and F.~Ringer,
Phys. Rev. D \textbf{102}, no.7, 074015 (2020)
doi:10.1103/PhysRevD.102.074015
[arXiv:2007.07281 [hep-ph]].

\bibitem{Kang:2020fka}
Z.~B.~Kang, X.~Liu, S.~Mantry and D.~Y.~Shao,
Phys. Rev. Lett. \textbf{125}, 242003 (2020)
doi:10.1103/PhysRevLett.125.242003
[arXiv:2008.00655 [hep-ph]].

\bibitem{Amilkar} A.~Quintero, EIC 2019 Users Group Annual Meeting, Paris, June 2019.

\bibitem{Miguel} M.~Arratia, DIS 2021, XXVIII International Workshop on Deep-Inelastic Scattering and Related Subjects, 2021.


\bibitem{Collins:1984kg}
J.~C.~Collins, D.~E.~Soper and G.~F.~Sterman,
Nucl. Phys. B \textbf{250}, 199-224 (1985)
doi:10.1016/0550-3213(85)90479-1

\bibitem{Su:2014wpa}
P.~Sun, J.~Isaacson, C.~P.~Yuan and F.~Yuan,
Int. J. Mod. Phys. A \textbf{33}, no.11, 1841006 (2018)
doi:10.1142/S0217751X18410063
[arXiv:1406.3073 [hep-ph]].

\bibitem{Prokudin:2015ysa}
A.~Prokudin, P.~Sun and F.~Yuan,
Phys. Lett. B \textbf{750}, 533-538 (2015)
doi:10.1016/j.physletb.2015.09.064
[arXiv:1505.05588 [hep-ph]].


\bibitem{Sun:2018icb} 
  P.~Sun, B.~Yan, C.-P.~Yuan and F.~Yuan,
  Phys.\ Rev.\ D {\bf 100}, no. 5, 054032 (2019)
  doi:10.1103/PhysRevD.100.054032
  [arXiv:1810.03804 [hep-ph]].
  
  \bibitem{Sun:2016kkh} 
  P.~Sun, J.~Isaacson, C.-P.~Yuan and F.~Yuan,
  Phys.\ Lett.\ B {\bf 769}, 57 (2017)
  doi:10.1016/j.physletb.2017.02.037
  [arXiv:1602.08133 [hep-ph]].
  
  
\bibitem{Liu:2020rvc}
T.~Liu, W.~Melnitchouk, J.~W.~Qiu and N.~Sato,
[arXiv:2008.02895 [hep-ph]].

\bibitem{Kidonakis:1998bk}
N.~Kidonakis, G.~Oderda and G.~F.~Sterman,
Nucl. Phys. B \textbf{525}, 299-332 (1998)
doi:10.1016/S0550-3213(98)00243-0
[arXiv:hep-ph/9801268 [hep-ph]].

\bibitem{Kidonakis:1998nf}
N.~Kidonakis, G.~Oderda and G.~F.~Sterman,
Nucl. Phys. B \textbf{531}, 365-402 (1998)
doi:10.1016/S0550-3213(98)00441-6
[arXiv:hep-ph/9803241 [hep-ph]].

\bibitem{Dominguez:2011br}
F.~Dominguez, J.~W.~Qiu, B.~W.~Xiao and F.~Yuan,
Phys. Rev. D \textbf{85}, 045003 (2012)
doi:10.1103/PhysRevD.85.045003
[arXiv:1109.6293 [hep-ph]].

\bibitem{Qiu:2011ai}
J.~W.~Qiu, M.~Schlegel and W.~Vogelsang,
Phys. Rev. Lett. \textbf{107}, 062001 (2011)
doi:10.1103/PhysRevLett.107.062001
[arXiv:1103.3861 [hep-ph]].

\bibitem{Boer:2011kf}
D.~Boer, W.~J.~den Dunnen, C.~Pisano, M.~Schlegel and W.~Vogelsang,
Phys. Rev. Lett. \textbf{108}, 032002 (2012)
doi:10.1103/PhysRevLett.108.032002
[arXiv:1109.1444 [hep-ph]].

\bibitem{Akcakaya:2012si}
E.~Akcakaya, A.~Sch\"afer and J.~Zhou,
Phys. Rev. D \textbf{87}, no.5, 054010 (2013)
doi:10.1103/PhysRevD.87.054010
[arXiv:1208.4965 [hep-ph]].

\bibitem{Pisano:2013cya}
C.~Pisano, D.~Boer, S.~J.~Brodsky, M.~G.~A.~Buffing and P.~J.~Mulders,
JHEP \textbf{10}, 024 (2013)
doi:10.1007/JHEP10(2013)024
[arXiv:1307.3417 [hep-ph]].

\bibitem{Boer:2013fca}
D.~Boer, W.~J.~den Dunnen, C.~Pisano and M.~Schlegel,
Phys. Rev. Lett. \textbf{111}, no.3, 032002 (2013)
doi:10.1103/PhysRevLett.111.032002
[arXiv:1304.2654 [hep-ph]].

\bibitem{Boer:2014lka}
D.~Boer and C.~Pisano,
Phys. Rev. D \textbf{91}, no.7, 074024 (2015)
doi:10.1103/PhysRevD.91.074024
[arXiv:1412.5556 [hep-ph]].


\bibitem{Boer:2020bbd}
D.~Boer, U.~D'Alesio, F.~Murgia, C.~Pisano and P.~Taels,
JHEP \textbf{09}, 040 (2020)
doi:10.1007/JHEP09(2020)040
[arXiv:2004.06740 [hep-ph]].



\bibitem{Sun:2011iw}
P.~Sun, B.~W.~Xiao and F.~Yuan,
Phys. Rev. D \textbf{84}, 094005 (2011)
doi:10.1103/PhysRevD.84.094005
[arXiv:1109.1354 [hep-ph]].

\bibitem{Gutierrez-Reyes:2019rug}
D.~Gutierrez-Reyes, S.~Leal-Gomez, I.~Scimemi and A.~Vladimirov,
JHEP \textbf{11}, 121 (2019)
doi:10.1007/JHEP11(2019)121
[arXiv:1907.03780 [hep-ph]].

\bibitem{Altinoluk:2021ygv}
T.~Altinoluk, C.~Marquet and P.~Taels,
[arXiv:2103.14495 [hep-ph]].

\bibitem{Boussarie:2021lkb}
R.~Boussarie, H.~M\"antysaari, F.~Salazar and B.~Schenke,
[arXiv:2106.11301 [hep-ph]].

\bibitem{Wang:2012xs}
J.~Wang, C.~S.~Li, Z.~Li, C.~P.~Yuan and H.~T.~Li,
Phys. Rev. D \textbf{86}, 094026 (2012)
doi:10.1103/PhysRevD.86.094026
[arXiv:1205.4311 [hep-ph]].

\bibitem{Boer:2014tka}
D.~Boer and W.~J.~den Dunnen,
Nucl. Phys. B \textbf{886}, 421-435 (2014)
doi:10.1016/j.nuclphysb.2014.07.006
[arXiv:1404.6753 [hep-ph]].

\bibitem{Collins:1981uw}
J.~C.~Collins and D.~E.~Soper,
Nucl. Phys. B \textbf{194}, 445-492 (1982)
doi:10.1016/0550-3213(82)90021-9

\bibitem{Mulders:2000sh}
P.~J.~Mulders and J.~Rodrigues,
Phys. Rev. D \textbf{63}, 094021 (2001)
doi:10.1103/PhysRevD.63.094021
[arXiv:hep-ph/0009343 [hep-ph]].

\bibitem{Ji:2005nu}
X.~d.~Ji, J.~P.~Ma and F.~Yuan,
JHEP \textbf{07}, 020 (2005)
doi:10.1088/1126-6708/2005/07/020
[arXiv:hep-ph/0503015 [hep-ph]].

\bibitem{Dominguez:2010xd}
F.~Dominguez, B.~W.~Xiao and F.~Yuan,
Phys. Rev. Lett. \textbf{106}, 022301 (2011)
doi:10.1103/PhysRevLett.106.022301
[arXiv:1009.2141 [hep-ph]].

\bibitem{Collins:2011zzd}
J.~Collins,
Camb. Monogr. Part. Phys. Nucl. Phys. Cosmol. \textbf{32}, 1-624 (2011)

\bibitem{Catani:2000vq}
S.~Catani, D.~de Florian and M.~Grazzini,
Nucl. Phys. B \textbf{596}, 299-312 (2001)
doi:10.1016/S0550-3213(00)00617-9
[arXiv:hep-ph/0008184 [hep-ph]].

\bibitem{Catani:2013tia}
S.~Catani, L.~Cieri, D.~de Florian, G.~Ferrera and M.~Grazzini,
Nucl. Phys. B \textbf{881}, 414-443 (2014)
doi:10.1016/j.nuclphysb.2014.02.011
[arXiv:1311.1654 [hep-ph]].

\bibitem{Xiao:2017yya}
B.~W.~Xiao, F.~Yuan and J.~Zhou,
Nucl. Phys. B \textbf{921}, 104-126 (2017)
doi:10.1016/j.nuclphysb.2017.05.012
[arXiv:1703.06163 [hep-ph]].

\end{thebibliography}

\end{document}